\documentclass{aa}  

\usepackage{graphicx}
\usepackage{txfonts}
\usepackage{soul}
\usepackage{xcolor}
\usepackage{multirow}
\usepackage{booktabs} 
\usepackage{tabularx} 
\usepackage{adjustbox} 
\usepackage{rotating}
\usepackage{lscape}
\usepackage{longtable}
\usepackage[flushleft]{threeparttable}
\usepackage{subfigure}
\usepackage{placeins}
\usepackage{hyperref}
\usepackage{hyperref}
\usepackage{tikz}
\usepackage[dvipsnames]{xcolor}

\usepackage{caption}
\DeclareCaptionFormat{cont}{#1 (cont.)#2#3\par}

\usepackage{ulem}
\begin{document}

   \title{Interferometric Survey of Stellar Parameters}

   \subtitle{Towards homogeneous FGK stars parameters and surface-brightness color relation in the context of PLATO space mission}

\author{Romina V. Ibañez Bustos  \inst{1}\fnmsep\thanks{\email{romina.ibanez@oca.eu}}
      \and Denis Mourard     \inst{1}
      \and Nicolas Nardetto \inst{1}
      \and Jeremy Jones \inst{2}
      \and Nayeem Ebrahimkutty       \inst{1}
      \and Juraj~Jonák       \inst{1}
      \and Hugo Nowacki     \inst{1}
      \and Mathieu Vrard     \inst{1}
      \and Philippe Berio       \inst{1}
      \and Julien Dejonghe       \inst{1}
      \and Roxanne Ligi      \inst{1}
      \and Frédéric Morand    \inst{1}
      \and David Salabert       \inst{1}
      \and Orlagh Creevey       \inst{1}
      \and Armando Domiciano de Souza  \inst{1}
      \and Manon Bailleul      \inst{1}
      \and Karine Perraut       \inst{3}
      \and Markus Wittkowski    \inst{4}
      \and John D. Monnier       \inst{5}
      \and Stefan Kraus         \inst{6}
      \and Narsireddy Anugu         \inst{2}
      \and Mayra Gutierrez   \inst{5}  
      \and Noura Ibrahim       \inst{5}
      }

\institute{Laboratoire Lagrange, Université Côte d’Azur, Observatoire de la Côte d’Azur, CNRS, Boulevard de l’Observatoire, CS 34229, 06304 Nice Cedex 4, France
     \and The CHARA Array of Georgia State University, Mount Wilson Observatory, Mount Wilson, CA 91203, USA
     \and Univ. Grenoble Alpes, CNRS, IPAG, 38000 Grenoble, France
     \and European Southern Observatory, Karl-Schwarzschild-Str. 2, 85748 Garching bei München, Germany
     \and Astronomy Department, University of Michigan, Ann Arbor, MI 48109, USA
     \and Astrophysics Group, Department of Physics \& Astronomy, University of Exeter, Stocker Road, Exeter, EX4 4QL, UK
     }

\date{Received 2025; accepted 2026}

 
  \abstract
   {The estimation of stellar angular diameters can be performed from the surface brightness -- color relation (SBCR) and photometry. The SBCRs have been considered by the PLATO space mission as an independent empirical alternative for estimating stellar radii of FGK stars.}
   {In this context, we have implemented an homogeneous approach not only for calibrating the SBCR for FGK-IV/V stars but, also for determining their fundamental parameters in order to place the stars reliably on the HR diagram and to study their impact on the  SBC relations. }
   {We have performed interferometric observations of 18 quiescent FGK-IV/V stars in the Gaia color range of \hbox{$3.088 \leq G \leq 5.498$}. For the first time, we used 3 different interferometric instruments operating in \textit{the R}, \textit{H} , and \textit{K} bands to measure polychromatic limb-darkened angular diameters. In parallel, by using public domain spectra we have estimated the stellar parameters ($T_\mathrm{eff}$, $\log g$ and Z) by using the open python tool \texttt{iSpec}.  }
   {We achieved an average accuracy of 2.3\% for the limb-darkened angular diameters based on polychromatic observations. However, we have observed that our SBCR relation does not follow the calibration of the relation between surface brightness and color in Gaia found in the literature. Furthermore, we found that $\log g$ and $Z$ have no impact on the SBC relationship; given the characteristics of our sample of quiescent stars, it constitutes an ideal set of targets for conducting a new SBCR calibration within the framework of the PLATO space mission. In this context, we reported a SBCR calibration with $\sigma_{RMS} = $ 0.012, 0.009, 0.009 in the $G, G_{BP}$ and $G_{RP}$ Gaia bands, respectively.  }
   {This article is part of a series of papers reporting the first results obtained using a polychromatic approach to measure angular diameters, employing a fully homogeneous methodology for both determining the fundamental parameters of stars and measuring the limb-darkened angular diameters, $\theta_{LD}$. }


   \maketitle
%

\newcommand{\RL}[1]{\textcolor{blue}{\bf{RL: #1}}}

\section{Introduction}

The surface brightness--colour relations (SBCRs) are key tools for easily estimating the angular diameter of a star from photometric measurements. 
They are of great interest for the study of stars hosting transiting exoplanets, since the planet's radius can then be derived directly from the stellar radius through its transit (e.g. \citealt{Ligi16}). 
An independent estimation of the stellar radius through the SBCRs derived with observations from the VEGA\footnote{Visible spEctroGraph and polArimeter (VEGA) was a visible instrument installed at the CHARA array \citep{Mourard2009}.} instrument, was preliminary implemented in the pipeline of the PLAnetary Transits and Oscillation of stars (PLATO) space mission \citep{Gent22}. 
This work is still under progress today.

The SBCRs are also widely used to derive the distances of eclipsing binaries in the Large and Small Magellanic Clouds (LMC and SMC, respectively) with high precision: 1\% for LMC (\citealt{Pietrzy13, Pietrzy19}) and 2\% for the SMC \citep{Graczyk20}. 
Additionally, the SBCRs are used to derive the distance of Cepheids (\citealt{Nardetto23, Bailleul2025}), or to estimate the angular diameter of interferometric calibrators.

Recent studies have shown that the SBCRs depend not only on the temperature of stars, but also on their luminosity class (\citealt{Salsi20, Salsi21}). 
This result was also confirmed from atmospheric models \citep{Salsi22}. 
Therefore, applying a homogeneous approach on a large sample of data is of critical importance to improve the precision and reliability of these relations. 

The Interferometric Survey of Stellar Parameters (ISSP) project was created with the aim of answering all these questions through an ambitious and homogeneous study of the angular diameters of a thousand bright stars (Vmag $\sim$ 8) and as small as 0.2 milliarcseconds (mas).
The project is based on the Stellar Parameters and Images with a Cophased Array (SPICA) instrument and has been designed to tackle fundamental questions about the relationship between planets and stars and to provide the wider community with a unique and primary source of direct information on a representative sample of stars across the HR diagram \citep{spica2026}. 
One of its core scientific programmes is the calibration of SBCRs with a typical precision of 1--2\%, depending on the stellar population. 
The programme aims to provide dedicated calibrations for late-type dwarfs and subgiants in support of PLATO, K giants and early-type stars for extragalactic distance determinations in the framework of Araucaria project, and to investigate the effects of spectral type, luminosity class, and stellar activity on the SBCR as we mentioned before.






To reach these goals and for operational aspects (multi-programme strategy of ISSP), we carefully selected an initial large set of 868 stars following the ideas and concepts from \cite{Salsi20, Salsi21} where the stars were identified with a very low activity level over all the HR diagram, i.e. classified as flag $=$ 0 in JSDC catalogue\footnote{The JMMC Stellar Diameter Catalog (JSDC): \url{https://www.jmmc.fr/english/tools/data-bases/jsdc-72/}.} and MAIN$\_$type $=$ STAR in SIMBAD to avoid biases in the interferometric angular diameter, but also in the photometry.
Among these stars, our goal is to observe 2 targets per sub-spectral channels [B0 - M3] and per class (V, IV, $>$III) which corresponds to 324 stars to be observed in the S04 ISSP survey.


In this article, we present the initial results of this programme through the analysis of a homogeneous sample of 18 quiet FGK dwarf and subgiant stars. 
This sample represents the first step towards calibrating the late-type SBCR required for the PLATO space mission and establishes the methodology to be applied to the full ISSP sample.
The paper is structured as follows:
in Section \ref{s.data} we describe the stellar sample, the interferometric and complementary data employed in this study. 
In Section \ref{s.analysis}, we describe the analysis carried out in this work for the obtention of the stellar parameters and surface brightness.
Then, in Section \ref{s.discussion} we discuss our results and we perform the calibration of the SBCR for the quiet stars studied here.
Finally, in Section \ref{s.conclusion} we highlight our conclusions of this work and we enumerate the steps to follow in order to achieve the rest of our goals.

\section{Data} \label{s.data}
\subsection{The stellar sample} \label{ss.sample}
\begin{table*}[htb!]
\caption{Stellar sample. We have sorted the target list by spectral type. The magnitudes reported here are from Gaia DR3 \citep{GaiaDR323} after the treatment carried out in section \ref{s.extinction} and the number of the spectra and the spectrograph implemented in this work are in columns 6 and 7, respectively. } 
\centering
\label{tab_sample}
\begin{tabular}{lcccccc}
\hline\hline\noalign{\smallskip}
Star      & SpT          & $G$                 & $G_{BP}$                & $G_{RP}$           & \# spectra & spectrograph \\
 & & [mag] & [mag] & [mag] & \\
\hline\noalign{\smallskip}
HD 173667 & F5.5IV-V     & 4.069 $\pm$ 0.037    & 4.306 $\pm$ 0.045    & 3.668 $\pm$ 0.029    & 86         & SOPHIE       \\
HD 142860 & F6V          & 3.694 $\pm$ 0.037    & 3.996 $\pm$ 0.048    & 3.293 $\pm$ 0.029    & 160        & HARPS        \\
HD 30652  & F6V          & 3.088 $\pm$ 0.037    & 3.514 $\pm$ 0.060    & 2.708 $\pm$ 0.032    & 39 / 56    & HARPS        \\
HD 4614   & F9V          & 3.320 $\pm$ 0.036    & 3.736 $\pm$ 0.053    & 2.849 $\pm$ 0.030    & 189        & SOPHIE       \\
HD 22484  & F9IV-V       & 4.134 $\pm$ 0.037    & 4.425 $\pm$ 0.044   & 3.671 $\pm$ 0.030   & 64         & SOPHIE       \\
HD 114710 & F9.5V        & \multicolumn{1}{l}{} & \multicolumn{1}{l}{} & \multicolumn{1}{l}{} & 19         & SOPHIE       \\
HD 19373  & G0V          & 3.903 $\pm$ 0.036    & 4.209 $\pm$ 0.044   & 3.427 $\pm$ 0.029    & 1          & SOPHIE       \\
HD 34411  & G1.5IV-VFe-1 & 4.520  $\pm$ 0.037   & 4.841 $\pm$ 0.044    & 4.053 $\pm$ 0.029    & 1          & ELODIE       \\
HD 182572 & G7IVHdel1    & 4.964 $\pm$ 0.036    & 5.341 $\pm$ 0.043    & 4.440 $\pm$ 0.029    & 22         & FEROS        \\
HD 190360 & G7IV-V       & 5.545 $\pm$ 0.036    & 5.913 $\pm$ 0.043    & 5.009 $\pm$ 0.029    & 24         & SOPHIE       \\
HD 10700  & G8V          & 3.300 $\pm$ 0.035    & 3.800 $\pm$ 0.051    & 2.748 $\pm$ 0.030    & 167 / 171  & HARPS        \\
HD 188512 & G8IV         & 3.482 $\pm$ 0.035    & 3.983 $\pm$ 0.049    & 2.873 $\pm$ 0.030    & 2060       & HARPS        \\
HD 185144 & K0V          & 4.449 $\pm$ 0.035    & 4.859 $\pm$ 0.043    & 3.865 $\pm$ 0.029   & 1278       & SOPHIE       \\
HD 3651   & K0.5V        & \multicolumn{1}{l}{} & \multicolumn{1}{l}{} & \multicolumn{1}{l}{} & 14         & HARPS-N      \\
HD 10476  & K1V          & 4.999 $\pm$ 0.035    & 5.440 $\pm$ 0.043    & 4.417 $\pm$ 0.029    & 2          & FEROS        \\
HD 4628   & K2.5V        & 5.468 $\pm$ 0.034    & 5.938 $\pm$ 0.042    & 4.829 $\pm$ 0.029    & 183        & HARPS        \\
HD 219134 & K3V          & 5.232 $\pm$ 0.034    & 5.767 $\pm$ 0.042    & 4.550 $\pm$ 0.028    & 341        & HARPS-N      \\
HD 16160  & K3V          & 5.498 $\pm$ 0.034    & 6.019 $\pm$ 0.042    & 4.821 $\pm$ 0.029    & 159 / 234  & HARPS   \\    
\hline\noalign{\smallskip}
\end{tabular}
\end{table*}


From the whole ISSP S04 target sample, 39 different stars have already been observed for the SBCR programme of ISSP project. 
Of these 39 stars, 10 targets have been observed only with the SPICA instrument and 26 stars have been observed twice or more times within the framework of the ISSP project. 

The sensitivity of SPICA is limited by the variable quality of the adaptive optics (AO) correction and the ageing of the coating of the telescopes. 
The current limiting magnitude in low resolution mode of SPICA is around $V=5.5$. 
On the other hand, the PLATO core sample focuses on dwarf and subgiant stars of spectral types ranging from F5 to K7 \citep{Montalto2021}.
Taking into account these two reasons, we have decided to study a sample of dwarfs and subgiants with spectral type ranging from F5.5 to K3 with visible magnitudes lower than 6 and estimated angular diameter greater than 0.6 mas, since our observational purpose is to acquire data from three different instruments working in different spectral bands (see Section \ref{ss.interferometry}).

The star sample studied in this paper is shown in Table \ref{tab_sample}.

\subsection{Interferometric observations from the ISSP project} \label{ss.interferometry}

\begin{table*}[htb!]
\caption{Observing log.}
\centering
\label{tab_observations}
\begin{tabular}{lcc|lcc|lcc}
\hline\hline\noalign{\smallskip}
Star      & Instrument & Obs.                                                                                      & Star      & Instrument & Obs.                                                                                      & Star      & Instrument & Obs.                                                                                                   \\
\hline\noalign{\smallskip}
HD 173667 & \begin{tabular}[c]{@{}c@{}} S \\ M\&M \\M\&M \\S\&M\&M\end{tabular}       & \begin{tabular}[c]{@{}c@{}}2023-08-09\\ 2024-06-12\\ 2025-04-23\\ 2025-05-09\end{tabular} & 
HD 19373  & \begin{tabular}[c]{@{}c@{}} MIRCX \\  S\&M\&M  \\ M\&M  \end{tabular}       & \begin{tabular}[c]{@{}c@{}}2023-10-17\\ 2024-10-14\\ 2025-10-07\end{tabular} & HD 185144 &  \begin{tabular}[c]{@{}c@{}} S\&M\&M \\  M\&M  \\ M\&M  \end{tabular}        & \begin{tabular}[c]{@{}c@{}}2025-06-15\\ 2025-08-14\\ 2025-08-15\end{tabular}                           \\
\hline
HD 142860 & \begin{tabular}[c]{@{}c@{}} S\&M\&M \\ S\&M\&M \\S\&M\&M \\ M\&M\end{tabular}         & \begin{tabular}[c]{@{}c@{}}2024-03-21\\ 2024-03-22\\ 2025-05-09\\ 2025-06-13\end{tabular} & HD 34411  & \begin{tabular}[c]{@{}c@{}} S\&M\&M \\  M\&M  \\ S\&M\&M  \\  M\&M \end{tabular}           & \begin{tabular}[c]{@{}c@{}}2024-10-31\\ 2025-09-13\\ 2025-10-07\\ 2025-12-13\end{tabular} & HD 3651   &\begin{tabular}[c]{@{}c@{}} M\&M \\ M\&M  \\ M\&M  \\  S\&M\&M \end{tabular}           & \begin{tabular}[c]{@{}c@{}}2025-08-13\\ 2025-08-18\\ 2025-09-14\\ 2025-10-05\end{tabular}              \\
\hline
HD 30652  & \begin{tabular}[c]{@{}c@{}} M\&M \\ S\&M\&M \\M\&M \\ S\&M\&M\end{tabular}         & \begin{tabular}[c]{@{}c@{}}2024-11-02\\ 2025-09-15\\ 2025-10-06\\ 2025-12-12\end{tabular} & HD 182572 & \begin{tabular}[c]{@{}c@{}} M\&M \\  MIRCX  \\ M\&M \end{tabular}            & \begin{tabular}[c]{@{}c@{}}2025-06-13\\ 2025-07-12\\ 2025-09-14\end{tabular}              & HD 10476  & \begin{tabular}[c]{@{}c@{}} S\&M\&M \\  M\&M  \\ M\&M  \\  S\&M\&M \end{tabular}        & \begin{tabular}[c]{@{}c@{}}2025-08-18\\ 2025-09-14\\ 2025-10-05\\ 2025-12-11\end{tabular}              \\
\hline
HD 4614   & \begin{tabular}[c]{@{}c@{}} M\&M \\ M\&M \end{tabular}           & \begin{tabular}[c]{@{}c@{}}2024-08-15\\ 2025-10-07\end{tabular}                           & HD 190360 & \begin{tabular}[c]{@{}c@{}} S\&M\&M \\ M\&M  \end{tabular}           & \begin{tabular}[c]{@{}c@{}}2025-07-14\\ 2025-08-18\end{tabular}                           & HD 4628   & \begin{tabular}[c]{@{}c@{}} M\&M \\  S\&M\&M  \\ M\&M  \end{tabular}         & \begin{tabular}[c]{@{}c@{}}2025-08-17\\ 2025-09-15\\ 2025-10-06\end{tabular}                           \\
\hline
HD 22484  & \begin{tabular}[c]{@{}c@{}} S\&M\&M \\ MIRCX \\M\&M \\ MIRCX\end{tabular}         & \begin{tabular}[c]{@{}c@{}}2023-11-12\\ 2024-11-02\\ 2025-11-13\\ 2025-12-12\end{tabular} & HD 10700  & \begin{tabular}[c]{@{}c@{}} M\&M \\  M\&M  \\ MIRCX \end{tabular}         & \begin{tabular}[c]{@{}c@{}}2024-08-18\\ 2024-09-14\\ 2025-10-06\end{tabular}              & HD 219134 & \begin{tabular}[c]{@{}c@{}} S\&M\&M \\ M\&M  \\ M\&M  \end{tabular}        & \begin{tabular}[c]{@{}c@{}}2024-07-06\\ 2025-08-14\\ 2025-08-15\end{tabular}                           \\
\hline
HD 114710 & \begin{tabular}[c]{@{}c@{}} S\&M\&M \\  S\&M\&M  \\ S\&M\&M  \\  S\&M\&M \end{tabular}         & \begin{tabular}[c]{@{}c@{}}2025-05-09\\ 2025-12-11\\ 2025-12-13\\ 2025-12-15\end{tabular} & HD 188512 & M\&M           & 2025-06-13                                                                                & HD 16160  & \begin{tabular}[c]{@{}c@{}} M\&M \\  S\&M\&M  \\ M\&M  \\  M\&M \\ M\&M \end{tabular}        & \begin{tabular}[c]{@{}c@{}}2025-08-17\\ 2025-09-15\\ 2025-10-06\\ 2025-11-13\\ 2025-12-12\end{tabular} \\
\hline
\end{tabular}
\tablefoot{\scriptsize{S: SPICA instrument; M\&M: MIRCX and MYSTIC instruments.}}
\end{table*}

We conducted our analysis using interferometric observations from three different instruments:
the SPICA instrument is composed of a visible spectrograph SPICA-VIS, a fibre-fed 6-beam interferometric spectrograph with three spectral resolutions, and SPICA-FT, a 6-beam near-infrared fringe tracker for fast stabilization of the fringes.
In addition, we also performed our observations by using the Michigan InfraRed Combiner-eXeter (MIRC-X) \citealt{Anugu2020}) instrument.
This instrument is a highly sensitive interferometer installed at the CHARA Array that provides an angular resolution equivalent of up to a 330 m diameter baseline telescope in J- and H-band wavelengths.
And finally, we carried out observations with the  Michigan Young STar Imager at CHARA (MYSTIC, \citealt{Monnier2018}). 
MYSTIC is a 6-beam combiner instrument operating in the K-band at CHARA.

We observed all the stars in our sample with the three instruments working simultaneously and using the 6--telescope configuration on the CHARA Array from August 2023 to December~2025. 
A summary of the observations is given in Table \ref{tab_observations}.
In the column ``Instrument'' we pointed out the data from the interferometers used for the final calibration of the data.
Our goal was to observe each target twice in order to avoid systematics. 
We note that most of the stars in the sample were observed on two or more occasions, with the exception of HD 188512, which was observed only once.
This is because SPICA was in the commissioning stage and it was decided to repeat the observations several times to improve the results. 
In the case of the star HD 188512, even if we observed it only once, we decided to include it as a complement to our work, but it will continue to be part of the sample to be observed during the current year. 

To perform these observations, we used the low spectral resolution mode of SPICA ($R \simeq 150$) for covering as much as possible the visible spectrum (600 - 900 nm, \citealt{spica2026}), a spectral dispersion of $R=50$ for MIRCX instrument required for sensitivity (\citealt{Anugu2020}) and $R = 100$ for MYSTIC for increasing the field-of-view (\citealt{Monnier2018}) for the fringe acquisition.
The integration time of our observations was of the order of 600 s on science targets and on each calibrator.

In order to calibrate the squared visibilities in spectral bands we have implemented the standard SPICA pipeline \citep{spica2026}.
The SPICA data reduction pipeline is implemented in \textsf{Python3.7} and carries out the following steps: first, the raw data is processed and spectral calibration is performed.
Next, it calculates the power spectrum of each spectral channel and estimates the spatial frequency of the centre of the fifteen fringe peaks.
To estimate the spatial frequency of each fringe, an artificial source called Six Telescope Simulator is used.
The squared visibilities for each fringe, the closure phase and the differential phase are then calculated using an image filtering based on the photometry of each telescope and the estimated SNR of the fringe.
The final step consists of calibrating the raw visibility, which means removing the instrumental signature by doing $V_{raw} = V_{inst}V_{exp}$ where $V_{inst}$ and $V_{exp}$ are the instrumental and expected visibilities, respectively.
The instrumental visibility or transfer function is computed as the ratio between the raw visibilities of the calibrators and their expected visibilities assuming a simple uniform disk model.
This pipeline produces science-ready data in OIFITS format \citep{Duvert2017} and is also adapted for MIRCX and MYSTIC data. 
In addition, for MIRCX and MYSTIC data, we performed a wavelength correction as suggested by the MIRCX/MYSTIC Pipeline Manual \footnote{See more details on v0.9.6, May 24 2024.} by dividing the measured wavelength by a constant factor (i.e. $\lambda_\mathrm{true} = \lambda_\mathrm{measured} / \kappa$): 
\begin{equation}
\begin{split}
    \kappa_\textrm{MIRCX} &= {1.0054 \pm 0.0006}, \\
    \kappa_\textrm{MIRCX} &= {0.999 \pm 0.001} \textrm{ (year 2025)}, \\
    \kappa_\textrm{MYSTIC} &= {1.0067 \pm 0.0007},\\
    \kappa_\textrm{MIRC} &= {1.0014 \pm 0.0006}.
\end{split}
\label{eq:mm_miscal}
\end{equation}

Regarding the calibrators, we have carefully selected each of them through the Night Scheduling Software (NSS, \citealt{spica2026}) that is connected to the JSDC catalog, taking into account non-variability, non-multiplicity, and that they are not resolved for SPICA, i.e., $V^2$ closest to unity, which allows for accurate measurement of the instrument's transfer function.
The calibrated files (L2 files) for each science target are available on the Optical interferometry DataBase (OiDB\footnote{\url{https://oidb.jmmc.fr}}). For each target, the list of calibrators with their respective angular diameters are contained in the header of the L2 files.

\subsection{Complementary data}

In this work, we have used publicly available photometry and spectroscopic observations to complement our different analyses. 
The programmes IDs of the spectroscopy data implemented in this work are shown in Appendix \ref{aa.complementData}.

\subsubsection{Photometry}

As demonstrated in \citealt{Salsi20, Salsi21}, SBCRs are highly dependent on photometry, and therefore accurate photometry is essential for their calibration \citep{Salsi22}.
Furthermore, homogeneity is a second important factor to consider when calibrating SBCRs.
Therefore, in this study we have taken into account the photometric bands G, G$_{BP}$, G$_{RP}$ from Gaia Data Release 3 (DR3, \citealt{GaiaDR323}), whose precision for our sample are 1.6\%, 1.3\% and 1\%, respectively (see columns 3, 4 and 5 of Table \ref{tab_sample}).

\subsubsection{Spectroscopy} \label{s.spectra}

In order to estimate the fundamental parameters of the stars (see Section \ref{fundParam}), we also complemented our data with public observations from different spectrographs:
\textit{HARPS}, mounted at the 3.6 m telescope at the European Southern Observatory (ESO, Chile). 
It covers a wavelength range from 380 to 690 nm with a resolving power of R = 115000 in 72 echelle orders \citep{Mayor2003}.

\textit{HARPS - North} is the Northern Hemisphere counterpart of HARPS. HARPS-N is installed at the Italian Telescopio Nazionale Galileo, located at the Roque de los Muchachos Observatory on the island of La Palma, Canary Islands, Spain \citep{Cosentino2012}.

\textit{FEROS}, placed on the 2.2 m telescope in ESO, is equipped with two fibres and operates in a wavelength range of 360-920 nm with a resolution of $R \sim 48000$ \citep{Kaufer99}.

\textit{SOPHIE}, is a high-resolution echelle spectrograph mounted on the 1.93 m reflector telescope at the Haute-Provence Observatory. Its wavelength coverage ranges from 387 to 690 nm and is capable of achieving resolutions of 40000 and 75000 in high efficiency  and high resolution modes, respectively \citep{Perruchot2008}.

\textit{ELODIE}, was an echelle spectrograph installed on the 1.93~m reflector at the Observatoire de Haute-Provence in south-eastern France. ELODIE was the ancestor of SOPHIE and its wavelength coverage was from 390 to 680 nm with a power resolution of R = 42000 \citep{Baranne1996}.

For our analysis we kept only those spectra with good signal-to-noise ratio (SNR) that we consider to be SNR $>$ 80.
In total, we have implemented 4809 spectra that have been automatically processed by their respective pipelines\footnote{\textsf{https://www.eso.org/sci/facilities/lasilla/instruments.html}}$^,$\footnote{\textsf{https://https://ohp.osupytheas.fr/sophie-echelle-spectrograph/}}$^,$\footnote{\textsf{http://atlas.obs-hp.fr/elodie/intro.html}}$^,$\footnote{\textsf{https://plone.unige.ch/HARPS-N}}. 
The number of spectra used for each star in our analysis is shown in column 6 of Table \ref{tab_sample}. 

\section{Analysis} \label{s.analysis}
In this section, we explain in detail the analysis performed for data from the stars HD 10700 ($\tau$ Cet) and HD 19373. The same procedure was followed for all the stars in our sample, except for a few differences that are explicitly discussed in the body of the text and/or in the discussions.
\subsection{Fundamental parameters} \label{fundParam}


In order to derive the fundamental parameters in a homogeneous way for all the stars in our sample, we use the open-source software \textsf{iSpec} \citep{ispec2019} for the analysis of the spectra employed in this work and described in section \ref{s.spectra}.

\textsf{iSpec} can derive fundamental parameters using the synthetic spectral-fitting technique and the equivalent-width (EW) method.
In this work, we have selected the method by measuring the equivalent-width of iron lines and, for the hottest stars in our sample, the synthetic spectral-fitting method. 
In the cool-stars regime, the spectra exhibit a high density of narrow, well-defined metallic lines, particularly those of Fe I and Fe II, which allows for precise EW measurements. 
Moreover, the assumptions of excitation and ionisation equilibrium are better satisfied, making the EW-based approach reliable and computationally efficient.

We used the MARCS grid\footnote{\textsf{https://marcs.astro.uu.se/}} \citep{Gustafsson2008}, solar abundances from \cite{Grevesse2007} and the line list from the Gaia-ESO survey \citep{Heiter2015}. 
Spectral regions contaminated with telluric lines were excluded from the analysis.
As input parameters, we implemented the values of $T_\mathrm{eff}$, $ \log g$ and $Z$ from Gaia DR3 and for those stars that are not part of Gaia DR3, we adopted the initial parameters from \cite{Soubiran2024}. 


For those targets with multiple spectra, we adopted the mean and standard deviation values as an estimation of the effective temperature, surface gravity, and their uncertainties, respectively.

In the case of HD 10700, 167 HARPS spectra with good SNR were implemented in this analysis.
We analysed each of them with iSpec and found the following fundamental parameters: \hbox{$T_\mathrm{eff} = 5382 \pm 52$ K,} $\log g = 4.38 \pm 0.03$ dex and $Z = -0.475 \pm 0.016$ dex.
For the case of the target HD 19373, only one spectrum was found in the public database of SOPHIE. 
In this special case, we reported the results obtained from  \texttt{iSpec} analysis with their respective uncertainties: $T_\mathrm{eff} = 5849 \pm 72$ K, $\log g = 4.05 \pm 0.09$ dex and $Z = 0.058 \pm 0.102$ dex.
For the rest of the stars in the sample, we report our results in the columns 2, 3 and 4 in the Table \ref{tab_resultsSpec} and the global comparison of our results with the literature can be found in the Appendix \ref{aa.literature}.

\subsection{Determination of $\theta_{LD}$} \label{s.angDiam}

For the determination of the limb-darkened angular diameters, we adopted a square-root law following the ideas from \cite{Ebrahimkutty2024}:

$$\frac{I(\mu)}{I(1)} = 1 - c(1- \mu) - d(1-\sqrt{\mu}),$$ where $I(1)$ is the specific intensity at the centre of the disc, $c$ and $d$ are the limb-darkening coefficients and $\mu$ is defined by $\mu = \cos{\gamma}$, where $\gamma$ is the angle between the line of sight and the emergent intensity.
We extracted the limb-darkening coefficients from \cite{Claret11} where they are tabulated based on ATLAS and PHOENIX plane-parallel atmosphere models. 
These tables require the fundamental parameters $T_\mathrm{eff}$, $\log g$ and $Z$ as inputs.
For the stars in our sample, we used the values found in this work in section \ref{fundParam} (see the columns 2, 3 and 4 in the Table \ref{tab_resultsSpec}).

We obtained the square visibilities for SPICA, MIRCX and MYSTIC observations by using the pipeline described in section \ref{ss.interferometry} (for more details, see \citealt{spica2026}).
From the measured visibilities, we derived $\theta_\textrm{LD}$ for each star in our sample by using the open tool \texttt{oimodeler}, a python-based optical-interferometric data modelling software \citep{Meilland2024}.
\texttt{oimodeler} allow us to build different models including chromaticity and time-dependence, among other.
As a starting point, we used the tables by \cite{Claret11} in the \textit{R}, \textit{H} and \textit{K} bands, since we observed in bands centered at 720, 1616 and 2170 nm, and we proceeded to perform interpolations in \texttt{oimodeler} to obtain a reliable coefficient at our wavelength.
We then estimated the angular diameters employing a model which considers a squared-root limb-darkening law and diffuse background flux. 
The diffuse background uniformly reduces the visibilities at all spatial frequencies, which is used to account for potential residual bias in the calibration process.
The uncertainties on angular diameters were derived from the $\chi^2$ profile by using $\chi^2_r(\theta_\textrm{fit}) = \text{min}(\chi^2_r)$ and $\chi^2_r(\theta_\textrm{fit}+\Delta\theta_{fit})=\text{min}(\chi^2_r) +1$ to account for underestimated observation errors and where $\theta_\textrm{fit}$ is the best solution of the fit.
We repeated the same procedure for each observed band, which means that for each star, we report the $\theta_\textrm{LD}$ for the three bands and the polychromatic angular diameter. 

\begin{figure}[htb!]
    \centering
    \includegraphics[width=0.45\textwidth]{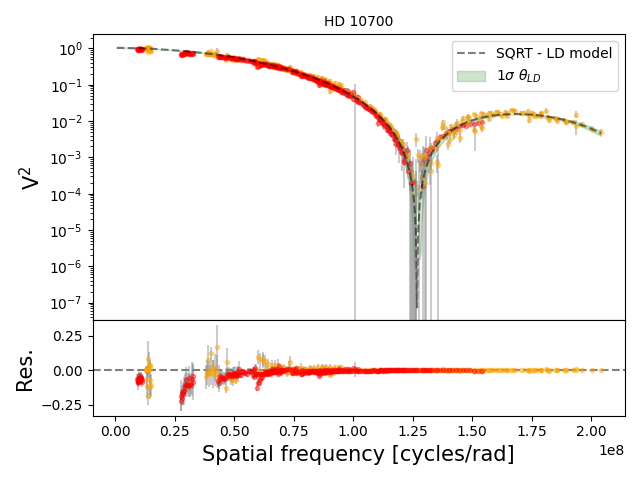}
    \includegraphics[width=0.45\textwidth]{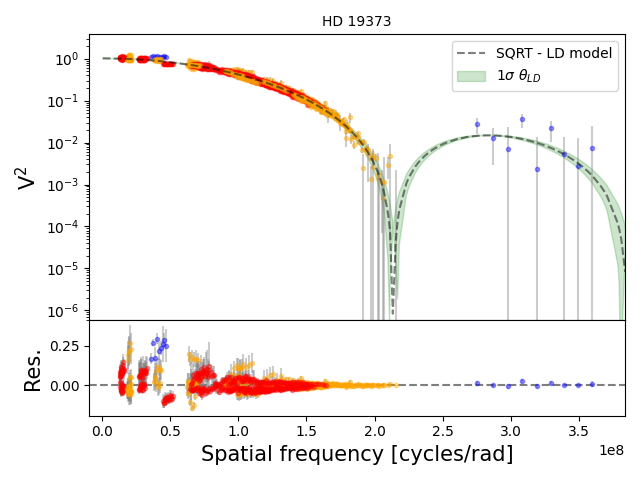}
    \caption{Squared visibilities of \textit{HD 10700} (top) \textit{HD 19373} (bottom). We represented SPICA, MIRCX and MYSTIC data in blue, orange and red points, respectively. The dashed lines represents the model of SQRT-LD angular diameter plus the background considered in this work and the green shadow represents the $1\sigma$ deviation.}
    \label{vis2_hd10700}
\end{figure}

At the top of Fig. \ref{vis2_hd10700} we show the MIRCX and MYSTIC data for the target HD 10700 observed during the nights 2024-08-18, 2024-09-14 and 2025-10-06. 
We fit the data with the model already mentioned where the dashed line represents the squared-root limb-darkening model and the green shadow is its 1$\sigma$ deviation.
For this target we measured an angular diameter $\theta_\textrm{SQRT-LD} = 2.042 \pm 0.021$ mas with a $\chi_r^2 = 1.597$ for the polychromatic data.
The angular diameters estimated in each band are reported in the Table \ref{tab_resultsInterf}.

For the case of HD 19373, we show the fitted data at the bottom of the Fig. \ref{vis2_hd10700}.
We observed this target four times during the nights 2023-10-17, 2024-09-15, 2024-10-14 and 2025-10-17 and we collected data from the 3 instruments.
From the \texttt{oimodeler} fit, we obtained $\theta_\textrm{SQRT-LD} = 1.217 \pm 0.016$ mas with a $\chi_r^2 = 2.976$ for the polychromatic data.
In both cases, we can see that the interferometric data is reaching the second lobe of the fitting.
This is fundamental if we want to extract more information from the star, i.e., the limb darkening coefficients as the second lobe of interferometric visibility is extremely sensitive to the intensity profile of the stellar disc. 
For HD 19373, only data in the \textit{R} band reach this regime, and it is particularly important for characterising smaller stars that fall outside the angular resolution provided by infrared instruments.
This is the first case where interferometric data in the visible range reach the second lobe.
The SPICA instrument offers a twofold advantage: it allows the second lobe to be explored for larger stars and the angular diameters of smaller stars to be determined with an angular resolution of up to 0.2 mas.

\begin{figure*}[htb!]
    \centering
    \begin{tikzpicture}
        \node[anchor=south west] (img) at (0,0) {\includegraphics[width=0.9\textwidth]{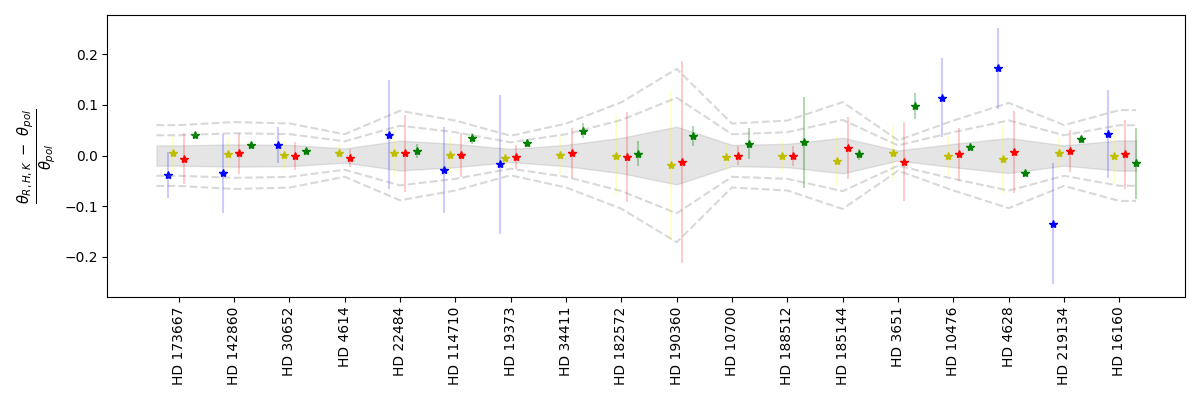}};
        \node[anchor=north west] at ([xshift=55pt,yshift=-12pt]img.north west) {\textcolor{blue}{\textbf{R}}, \textcolor{GreenYellow}{\textbf{H}}, \textcolor{red}{\textbf{K}} bands};
        \node[anchor=north west] at ([xshift=55pt,yshift=-22pt]img.north west) {\textcolor{OliveGreen}{\textbf{JMDC}}};
    \end{tikzpicture}

    \caption{Angular diameters measured for each star  normalised  to the  polychromatic $\theta_{LD}$. The gray shadow and the gray dashed lines corresponds to 1,2,3$\sigma$ of polychromatic $\theta_{LD}$, respectively. We also plot in green the angular diameters found in the literature in the JMDC catalog. }
    \label{angDiamNorm}
\end{figure*}

In Fig. \ref{angDiamNorm}, we show the angular diameters measured for each star normalised to the polychromatic $\theta_\textrm{SQRT-LD}$.
The green symbols show how our derived diameters compare to those reported in the JMDC catalogue\footnote{The JMMC Measured stellar Diameters Catalog:  \href{https://vizier.cds.unistra.fr/viz-bin/VizieR?-source=II/345}{II/345/jmdc}.} (included as part of \citealt{Chelli2016}).
We can see that the angular diameter measurements in \textit{R} band (in blue) for the targets HD 10476, HD 4628 and HD 219134 deviate from the polychromatic $\theta_\textrm{LD}$ but within its 3$\sigma$.
Those targets have only one observation carried out with SPICA instrument and in the squared visibilities shown in the Appendix \ref{aa.vis}, we can see that in \textit{R} band, the data are more spread and some baselines are not well represented by the model fit.
We need to continue observing these targets with SPICA instrument in order to improve the current situation.

For the whole sample, we have obtained an average accuracy of 2.3\% in angular diameters, which is consistent with the requirements established when selecting the targets with the main purpose of calibrating the SBCR.

\subsection{Stellar radii, $F_{bol}$ and $L\star$} \label{ss.radiiLum}

We estimated the stellar radii by combining the $\theta_{LD}$ obtained in \ref{s.angDiam} and the parallaxes, $\Pi$, from Gaia DR3:
\begin{equation}
    R_{\star} = \frac{\theta_{LD}}{9.301 ~\times ~ \Pi},
    \label{radii}
\end{equation}
where the constant term is based on the most recent solar constants \citep{Prsa16}.

We also estimated the bolometric flux, $F_{bol}$, taking into account the stellar angular diameters and their effective temperatures obtained in this work following
\begin{equation}
    F_{\textrm{bol} \star} = \frac{\sigma_\textrm{SB}}{4}  ~ \theta_\textrm{LD}^2 \times T_{\textrm{eff} \star}^{4},
    \label{fbol}
\end{equation}
where $\sigma_\textrm{SB}$ is the Stefan-Boltzmann constant.

Finally, we calculated the stellar luminosity $L_\star$ by combining the bolometric flux and
the parallax:
\begin{equation}
    L_{\star} = 4\pi d^2 F_{\textrm{bol} \star},
    \label{lum}
\end{equation}
where we considered $d=1/\Pi$. 

To determine the uncertainties in Eqs. \ref{radii}, \ref{fbol} and \ref{lum}, we considered the parameters to the right of each equation to be independent random variables with Gaussian probability density functions. 
Then, the standard deviation of each parameter is obtained analytically in the first order by applying a classical error propagation:
\begin{equation}
\begin{split}
    \bar{\sigma}_R{_\star} & = \sqrt{\bar{\sigma}_{\theta_{LD}}^2 + \bar{\sigma}_\Pi^2} \\ 
    \bar{\sigma}_{F_{bol}} & = \sqrt{ ~ (2 \times \bar{\sigma}_{\theta_{LD}})^2 + (4\times \bar{\sigma}_{T_\mathrm{eff}})^2} \\
    \bar{\sigma}_L{_\star} & = \sqrt{\bar{\sigma}_{F_{bol}}^2 + 4\times\bar{\sigma}_\Pi^2} 
    \label{sigma_radii},
\end{split}
\end{equation}
where $\bar{\sigma_i} = \sigma_i / i$ and $i$ refers to the different parameters referred to in the equations. $\sigma_{\theta_{LD}}$, $\sigma_\Pi$ and $\sigma_{T_\mathrm{eff}}$ are the uncertainties in LD angular diameter, parallax and effective temperature, respectively.
Our results are shown in the columns 5, 6 and 7 of Table \ref{tab_resultsSpec}.

\subsection{Extinction} \label{s.extinction}

We estimated the extinction thanks to the recent 3D maps that were provided by \cite{Vergely2022}.
These maps are based on the tomographic inversion following a hierarchical technique of two different extinction datasets, one combining spectroscopy and photometry and the second one, based on photometry of Gaia eDR3 and 2MASS.

Due to the proximity of the stars in our sample, we have chosen the map covering a volume of 3 kpc x 3 kpc x 0.8 kpc around the Sun, which is also the one with the highest resolution (10 pc).
The stars in our sample span distances from 4 pc to 20 pc, which were computed from their parallaxes taken from Gaia DR3.
Therefore, the estimated extinction is very low for most stars, ranging from 0.007 to 0.01 magnitudes.

For the Gaia band, we have implemented the formula from \cite{Danielski18}:
\begin{align*}
k_m =& ~ a_1 + a_2X + a_3X^2 + a_4X^3 + a_5A_V + a_6A_V^2 + a_7A_V^3  \\ 
&+ a_8A_VX + a_9A_VX^2 + a_{10}XA_V^2,
\end{align*}
where $k_m = A_m/A_V$ and $m = G$, $G_{BP}$ and $G_{RP}$.
In our case, the coefficients $a$ were computed for main sequence stars and we adopted $X = T_\mathrm{eff}/5040$ K, where we considered $T_\mathrm{eff}$ as the estimated temperature computed in our work.

Finally,  the Gaia magnitudes of the stars in the sample were corrected to account for extinction, and their uncertainties were obtained by propagating the uncertainties in the input quantities used to calculate the extinction coefficients.

\subsection{Surface brightness} \label{ss.sb}

\begin{figure}[htb!]
    \centering
    \includegraphics[width=0.42\textwidth]{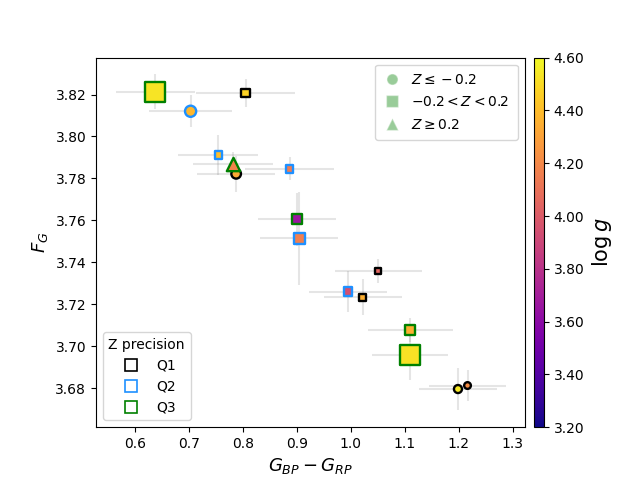}
    \includegraphics[width=0.40\textwidth]{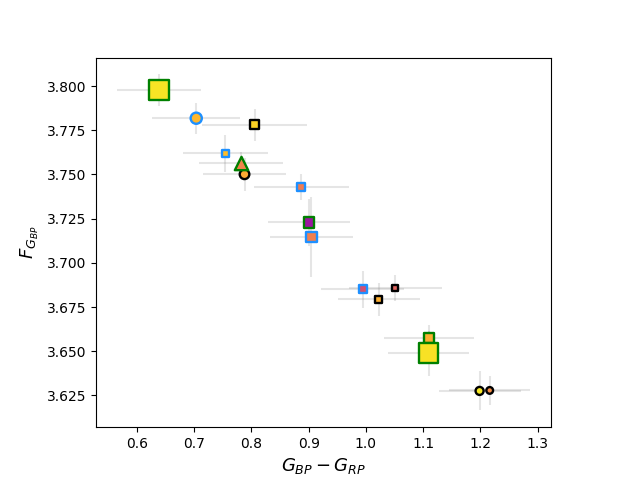}
    \includegraphics[width=0.40\textwidth]{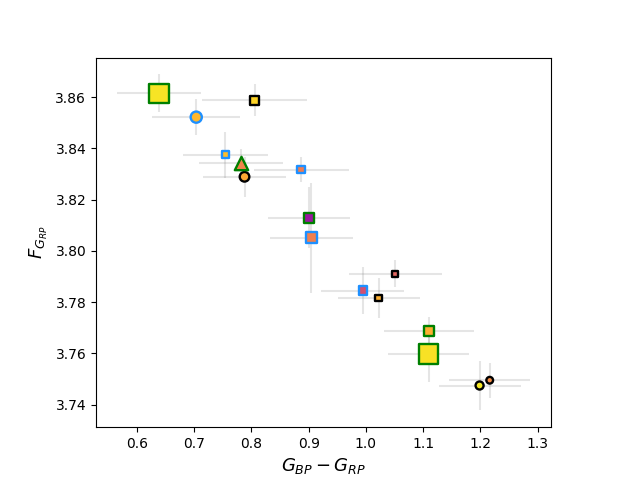}
    \caption{Surface brightness, $F_G$, ${F_G}_{BP}$ and ${F_G}_{RP}$ as a function of Gaia color $(G_{BP} - G_{RP})$. Marker color shows $\log g$, marker size scales with the relative uncertainty in $\log g$ (larger markers correspond to more precise measurements), shape indicates metallicity and edge color encodes metallicity precision.}
    \label{sbcr_logg_z}
\end{figure}

The surface brightness (SB) of a star, $S_\lambda$, is the flux density emitted per unit of angular area and is correlated to its limb-darkening angular diameter $\theta_\mathrm{LD}$ by the following formula:
\begin{equation}
    S_\lambda = m_{\lambda0} + 5\log \theta_\mathrm{LD},
    \label{S_theta}
\end{equation}
where $m_{\lambda0}$ is the apparent magnitude corrected from the extinction.

The definition of the SBCR is also coming from \cite{Wesselink69} who used the $S_\lambda$ definition to show the correlation between the surface brightness and colour of a given star as:
\begin{equation}
    S_{\lambda1} =  \sum_{n=0}^{N} C_n (m_{\lambda1} - m_{\lambda2})_0^n,
    \label{S_coeff}
\end{equation}
where $C_n$ is a constant that depends on the solar parameters (bolometric magnitude, $M_\mathrm{bol}$, and its total flux, $f_\odot$).

Later, \cite{Barnes76} built a linear relation between the surface brightness, noted as $F_\lambda$, and the stellar colour that can be expressed as follows:
\begin{equation}
    F_{\lambda1} = 4.2196 - 0.1 \times  \sum_{n=0}^{N} C_n (m_{\lambda1} - m_{\lambda2})_0^n,
    \label{F_coeff}
\end{equation}
where the value 4.2196 can be obtained from the formula extracted from \cite{Fouque1997}, taking into account the Stefan-Boltzmann constant, $\sigma_\mathrm{SB}$, $M_\mathrm{bol}$ and $f_\odot$ from \cite{Mamajek15} and \cite{Prsa16}.

Therefore, from the equations presented here, we can calibrate the SBCR by measuring the angular diameters of stars using interferometric observations and, conversely, estimate the stellar angular diameters for those targets that cannot be observed using interferometry by using SBC relations.

In Fig. \ref{sbcr_logg_z} we show the surface brightness $F_G$ (top), ${F_G}_{BP}$ (center) and ${F_G}_{RP}$ (bottom) as a function of Gaia color $(G_{BP} - G_{RP})$, where the SB was estimated through equations \ref{S_theta} and \ref{F_coeff} by using the angular diameters reported in section \ref{s.angDiam}.
In our plot, marker colours show $\log g$ and their size scales are the relative uncertainty in $\log g$, where larger markers correspond to more precise measurements.
The shape of each marker indicates metallicity and edge colour encodes metallicity precision.
By analysing $F_\lambda$, as a function of $\log g, ~Z$, we estimated the Pearson coefficients to evaluate if there is any correlation between the variables.
Performing an orthogonal distance regression with errors in both variables, we found the following Pearson or $R$-coefficients:

\[
  R =
  \left[ {\begin{array}{ccc}
     & \log g & Z \\
    F_G & -0.350 & 0.013 \\
    {F_G}_{BP} & -0.368 & 0.009 \\
    {F_G}_{RP} & -0.355 & 0.013 
  \end{array} } \right]
\]
In this work, we did not find any relation between the derived SB, $F_\lambda$, and $\log g, ~Z$.
The lack of dependence on metallicity can be explained by the fact that $Z$ found for our stars varies within a very small range between -0.478 and 0.339 dex (see Table \ref{tab_resultsSpec}).
It is essential to consider stars with different metallicities ranging between -2 and 1 dex, as suggested in the theoretical study by \cite{Salsi22} if one wants to further explore the analysis of the dependence of the SBCR on metallicity based on Gaia.
As for surface gravity, we can see at a glance that there is no dependence in the SBC graphs shown in Fig. \ref{sbcr_logg_z}.
This is expected given that our sample only contains FGK dwarfs and subgiants.
A more exhaustive analysis of how the difference between giant stars and dwarfs/subgiants may or may not impact or vary the SBC relationship is proposed for the other objectives of the ISSP-S04 project mentioned in the introduction to this work.



\begin{table*}[htb!]
\centering
\caption{Stellar parameters calculated from spectroscopy ($T_\mathrm{eff}$, $\log g$ and $Z$ ) and from a combination between spectroscopy and interferometry ($R_\star$, $F_{bol}$ and $L_\star [L_\odot]$).}
\label{tab_resultsSpec}
\resizebox{\textwidth}{!}{%
\begin{tabular}{lcccccc}
\hline\hline\noalign{\smallskip}
Star      & $T_\mathrm{eff} [K]$     & $\log g$        & $Z$                & $R_\star [R_\odot]$            & $F_{bol} [\times 10^{+8}]$ & $L_\star [L_\odot]$            \\
\hline\noalign{\smallskip}
HD 173667 & 6303 $\pm$ 21 & 3.93 $\pm$ 0.02 & -0.070 $\pm$ 0.020 & 2.005 $\pm$ 0.042    & 48.6 $\pm$ 2.1    & 5.68 $\pm$ 0.25    \\
HD 142860 & 6306 $\pm$ 37 & 4.15 $\pm$ 0.06 & -0.127 $\pm$ 0.016 & 1.437 $\pm$ 0.025    & 74.8 $\pm$ 3.3      & 2.92 $\pm$ 0.13    \\
HD 30652  & 6549 $\pm$ 17 & 4.33 $\pm$ 0.02   & -0.016 $\pm$ 0.010      & 1.311 $\pm$ 0.018    & 140.5 $\pm$ 4.2     & 2.83 $\pm$ 0.08   \\
HD 4614   & 5879 $\pm$ 27 & 4.22 $\pm$ 0.08 & -0.381 $\pm$ 0.028  & 1.036 $\pm$ 0.011    & 102.8 $\pm$ 2.6    & 1.13 $\pm$ 0.03    \\
HD 22484  & 5917 $\pm$ 18 & 3.94 $\pm$ 0.05 & -0.107 $\pm$ 0.017 & 1.609 $\pm$ 0.042    & 46.9 $\pm$ 2.6     & 2.84 $\pm$ 0.16    \\
HD 114710 & 6038 $\pm$ 23 & 4.47 $\pm$ 0.05 & 0.087 $\pm$ 0.022  & 1.079 $\pm$ 0.022    & 52.6 $\pm$ 2.4      & 1.39 $\pm$ 0.06    \\
HD 19373  & 5849 $\pm$ 72 & 4.05 $\pm$ 0.09 & 0.058 $\pm$ 0.102  & 1.383 $\pm$ 0.018    & 57.6 $\pm$ 3.1      & 2.02 $\pm$ 0.11    \\
HD 34411  & 5936 $\pm$ 65 & 4.35 $\pm$ 0.08 & 0.113 $\pm$ 0.086  & 1.261 $\pm$ 0.027    & 36.1 $\pm$ 2.3      & 1.78 $\pm$ 0.11    \\
HD 182572 & 5617 $\pm$ 8  & 4.18 $\pm$ 0.02 & 0.339 $\pm$ 0.005  & 1.348 $\pm$ 0.032    & 23.5 $\pm$ 2.0      & 1.64 $\pm$ 0.14    \\
HD 190360 & 5536 $\pm$ 17 & 4.16 $\pm$ 0.03 & 0.122 $\pm$ 0.014  & 1.156 $\pm$ 0.058    & 14.1 $\pm$ 2.4      & 1.13 $\pm$ 0.19    \\
HD 10700  & 5382 $\pm$ 52  & 4.38 $\pm$ 0.03 & -0.475 $\pm$ 0.016 & 0.801 $\pm$ 0.008    & 115.2 $\pm$ 5.1     & 0.48 $\pm$ 0.02    \\
HD 188512 & 5163 $\pm$ 47  & 3.66 $\pm$ 0.03 & -0.124 $\pm$ 0.008 & 3.123 $\pm$ 0.033    & 106.9 $\pm$ 4.5     & 6.18 $\pm$ 0.26    \\
HD 185144 & 5336 $\pm$ 40 & 4.33 $\pm$ 0.04 & -0.322 $\pm$ 0.037 & 0.776 $\pm$ 0.020    & 42.3 $\pm$ 2.7      & 0.44 $\pm$ 0.03    \\
HD 3651   & 5267 $\pm$ 13 & 4.52 $\pm$ 0.01 & 0.170 $\pm$ 0.006  & 0.864 $\pm$ 0.039    & 13.3 $\pm$ 0.4      & 0.51 $\pm$ 0.02   \\
HD 10476  & 5221 $\pm$ 4  & 4.53 $\pm$ 0.01 & 0.003 $\pm$ 0.001  & 0.808 $\pm$ 0.016    & 23.9 $\pm$ 1.1      & 0.44 $\pm$ 0.02    \\
HD 4628   & 4991 $\pm$ 61 & 4.54 $\pm$ 0.06 & -0.310 $\pm$ 0.022 & 0.721 $\pm$ 0.027    & 16.7 $\pm$ 1.5      & 0.29 $\pm$ 0.03   \\
HD 219134 & 4880 $\pm$ 46 & 4.35 $\pm$ 0.04 & 0.140 $\pm$ 0.010  & 0.756 $\pm$ 0.017    & 21.7 $\pm$ 1.2      & 0.29 $\pm$ 0.02    \\
HD 16160  & 4798 $\pm$ 42 & 4.40 $\pm$ 0.07 & -0.160 $\pm$ 0.011 & 0.743 $\pm$ 0.024    & 16.1 $\pm$ 1.2      & 0.26 $\pm$ 0.02    \\
\hline\noalign{\smallskip}

\end{tabular}
}
\tablefoot{$T_\mathrm{eff}$, $\log g$ and $Z$ were obtained through a Gaussian fit analysis for the stars highlighted with `$\star$'.}
\end{table*}

\begin{table*}[htb!]
\centering
\caption{Angular diameters of our targets (in mas) for each band $R, H, K$ and the polychromatic ones ($\theta_{LD}$)}
\label{tab_resultsInterf}
\resizebox{\textwidth}{!}{%
\begin{tabular}{lcccccccc}
\hline\hline\noalign{\smallskip}
Star      & $\theta_R$           & $\chi_r^2$              & $\theta_H$           & $\chi_r^2$              & $\theta_K$           & $\chi_r^2$              & $\theta_{LD}$             & $\chi_r^2$              \\
\hline\noalign{\smallskip}
HD 173667 & 0.924 $\pm$ 0.023 & 51.34  & 0.966 $\pm$ 0.012 & 0.73  & 0.955 $\pm$ 0.028 & 0.74  & 0.961 $\pm$ 0.020 & 0.87  \\
HD 142860 & 1.15 $\pm$ 0.07   & 1.64   & 1.195 $\pm$ 0.016 & 1.71  & 1.197 $\pm$ 0.026 & 0.95   & 1.192 $\pm$ 0.022 & 1.35  \\
HD 30652  & 1.544 $\pm$ 0.033 & 2.42   & 1.514 $\pm$ 0.021 & 2.99  & 1.514 $\pm$ 0.021 & 1.04  & 1.514 $\pm$ 0.021 & 1.64   \\
HD 4614   & --                & --      & 1.615 $\pm$ 0.013 & 6.45  & 1.599 $\pm$ 0.014 & 10.95 & 1.607 $\pm$ 0.014 & 12.13  \\
HD 22484  & 1.116 $\pm$ 0.087 & 0.12   & 1.077 $\pm$ 0.021 & 1.01   & 1.076 $\pm$ 0.052 & 0.26   & 1.072 $\pm$ 0.029 & 0.71  \\
HD 114710 & 1.059 $\pm$ 0.069 & 1.51   & 1.091 $\pm$ 0.016 & 1.71   & 1.091 $\pm$ 0.025 & 1.05   & 1.090 $\pm$ 0.023 & 1.23   \\
HD 19373  & 1.195 $\pm$ 0.154 & 1.07   & 1.210 $\pm$ 0.012 & 12.07  & 1.212 $\pm$ 0.016 & 6.51   & 1.216 $\pm$ 0.013 & 9.35  \\
HD 34411  & --                & --      & 0.936 $\pm$ 0.016 & 2.31  & 0.364 $\pm$ 0.025 & 1.75   & 0.935 $\pm$ 0.021 & 2.01   \\
HD 182572 & --                & --      & 0.842 $\pm$ 0.032 &  2.28  & 0.842 $\pm$ 0.040 & 0.86    & 0.842 $\pm$ 0.035 & 1.50  \\
HD 190360 & --                & --      & 0.659 $\pm$ 0.041 & 1.39   & 0.659 $\pm$ 0.076 & 0.63   & 0.672 $\pm$ 0.057 & 1.07  \\
HD 10700  & --                & --      & 2.024 $\pm$ 0.031 & 1.85   & 2.030 $\pm$ 0.018 & 1.12   & 2.030 $\pm$ 0.021 & 1.64   \\
HD 188512 & --                & --      & 2.123 $\pm$ 0.044 & 0.72   & 2.123 $\pm$ 0.020 & 0.36   & 2.125 $\pm$ 0.023 & 0.61   \\
HD 185144 & --                & --      & 1.238 $\pm$ 0.027 & 2.18   & 1.269 $\pm$ 0.041 & 1.61   & 1.251 $\pm$ 0.035 & 1.92   \\
HD 3651   & --                & --      & 0.724 $\pm$ 0.030 & 2.10  & 0.711 $\pm$ 0.046 & 1.17  & 0.720 $\pm$ 0.010 & 1.53   \\
HD 10476  & 1.095 $\pm$ 0.056 & 1.06   & 0.982 $\pm$ 0.016 & 1.20   & 0.985 $\pm$ 0.028 & 1.12    & 0.983 $\pm$ 0.023 & 1.12    \\
HD 4628   & 1.054 $\pm$ 0.043 & 5.73   & 0.892 $\pm$ 0.026 & 1.95   & 0.905 $\pm$ 0.039 & 1.32   & 0.899 $\pm$ 0.035 & 2.04   \\
HD 219134 & 0.928 $\pm$ 0.105 & 1.17   & 1.077 $\pm$ 0.013 & 1.27  & 1.081 $\pm$ 0.025 & 0.77   & 1.072 $\pm$ 0.020 & 1.14   \\
HD 16160  & 0.996 $\pm$ 0.054 & 1.31   & 0.954 $\pm$ 0.022 & 1.54   & 0.958 $\pm$ 0.035 & 1.34   & 0.956 $\pm$ 0.030 & 1.47                 \\
\hline\noalign{\smallskip}
\end{tabular}
}
\end{table*}

\section{Discussions} \label{s.discussion}

In this work, we presented for the first time a multichromatic analysis for determining the angular diameters of the stars in our sample.
Furthermore, this is the one of the first time that an interferometer operating in the optical range has reached the second lobe of the visibility curves.
This is fundamental not only for measuring limb darkening coefficients, which are critical for the study of exoplanets, but also for measuring the angular diameters of smaller stars, which until now had been a limitation.

It should be noted that these are the first results obtained with the SPICA instrument among another batch of papers and that the performance of the instrument is still being optimised.
As shown in Fig. \ref{angDiamNorm}, many angular diameters could be measured with this instrument; however, repeated observations of SPICA remain essential.

In the following sub-sections, we discuss about the points we considered most relevant according to our procedure for obtaining results (see Section \ref{s.analysis}).

\subsection{Fundamental parameters}

Following an approach that was as homogeneous as possible, we proceeded to obtain the fundamental parameters of the stars in our sample using public domain spectra in the different spectroscopic databases.

\begin{figure}[htb!]
    \centering
    \includegraphics[width=0.45\textwidth]{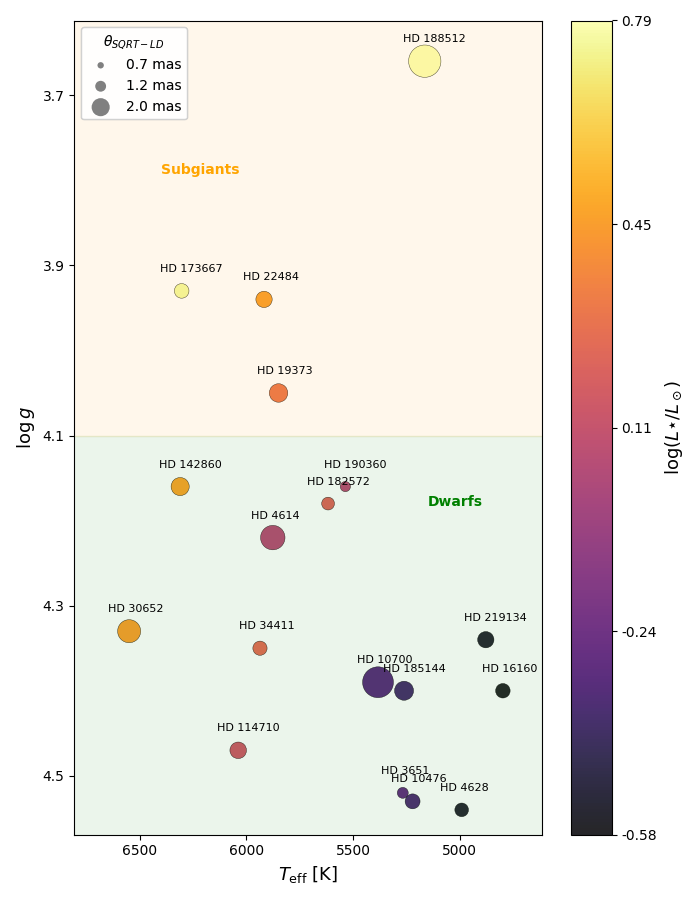}
    \caption{$\log g$ vs. $T_\mathrm{eff}$ diagram for the stars in our sample. The colour map show the $L_{\star}/L_{\odot}$ levels found in section \ref{ss.radiiLum} and the size of each circle represent de $\theta_{LD}$ measured in section \ref{s.angDiam}. We also divided the dwarf and subgiant regimes with green and yellow shading for better distinction in the characterisation of the stars in our sample.   }
    \label{diagramaHRres}
\end{figure}

In Fig. \ref{diagramaHRres} we show the position of the stars studied in the $\log g$ vs. $T_\mathrm{eff}$ diagram built with the fundamental parameters obtained in this work.
The color map correspond to $L\star / L_\odot$ and the size of the circles represents the angular diameters measured in this work.

Thanks to this analysis, we were able to place four stars studied reliably in the $\log g$ vs. $T_\mathrm{eff}$ diagram, which was one of our main objectives highlighted in the introduction to this work.
The target HD 19373 was considered as a dwarf in Simbad, and in our analysis falls within the subgiant range. 
Furthermore, for the cases of HD 34411, HD 182572 and HD 190360, they were considered to be in the subgiant-dwarf transition, but in this work they are clearly dwarf stars. 

We were also able to estimate the stellar radii for the stars in our sample with a precision between 1\% and 5\%, taking into account the angular diameters measured in this work and the parallaxes from Gaia DR3. 
Furthermore, through $\theta_\mathrm{LD}$, $T_\mathrm{eff}$, and $F_\mathrm{bol}$, we could calculate the values of stellar luminosities, \hbox{$L_\star/L_\odot$}, with an accuracy of between 2 and 10\%.

\begin{figure}[htb!]
    \centering
    \includegraphics[width=0.45\textwidth]{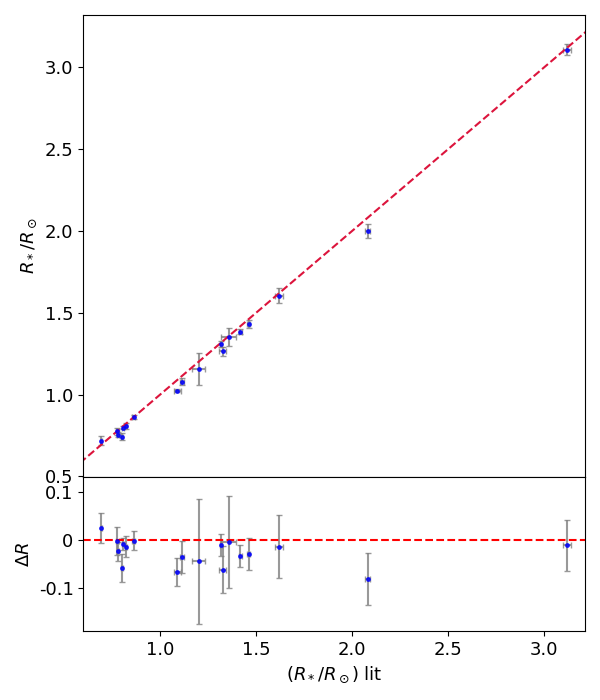}
    \includegraphics[width=0.45\textwidth]{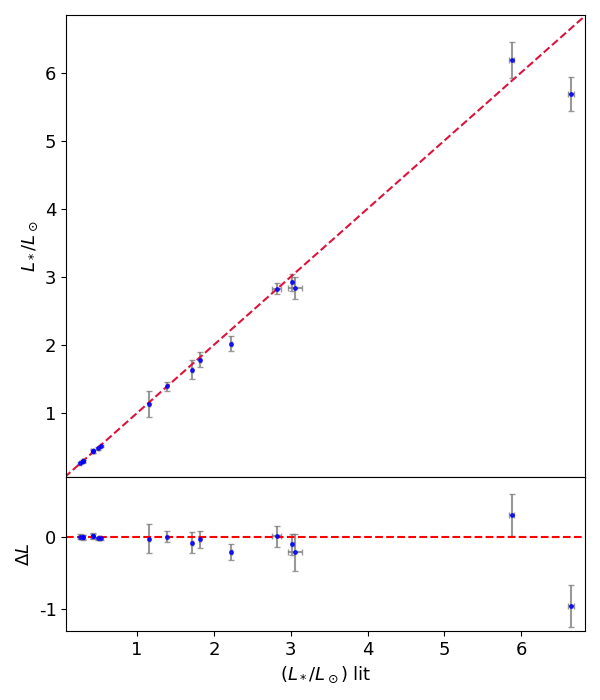}
    \caption{Stellar radii (top) and $L\star / L_\odot$ (bottom). Comparison between what we found in the literature (mainly from \citealt{Hirsch2021, Soubiran2024}) and the values calculated in this work. In each graph, we also show the difference $\Delta X = X - X \rm ~ lit$, where $X = R_\star / R_\odot$, $L_\star / L_\odot$. Red dashed lines provides 1:1 relations.  }
    \label{compLit}
\end{figure}

In many studies found in the literature, $L_\star/L_\odot$ is obtained through an SED-type fit based on photometric data and spectral libraries covering a wide wavelength range.
In such cases, $T_\mathrm{eff}$, $\log g$, $Z$, $A_V$, and a scaling factor that define the characteristics of the spectra are generally required as input parameters (see \cite{Ligi16} for example). 
In this type of study, uncertainties usually arise from the fitting.
In our case, we use observed stellar spectra to estimate $T_\mathrm{eff}$, $\log g$ and $Z$, using atmospheric models and atomic lines as described in section \ref{fundParam}.
Our uncertainties derive from a statistical analysis of several processed spectra.
Finally, this analysis allows us to derive the rest of the parameters, such as $F_\mathrm{bol}$ and $L_\star/L_\odot$.
In both cases, we have to pay the price of using models to carry out the fitting. 
In Fig. \ref{compLit}, we show the results of the stellar radii and $L\star/L_\odot$ obtained in this study and the comparison with the values found in the literature mainly from \cite{Soubiran2024}.

We can see that our results are in good agreement with those found in the literature.
For the case of $L_\star/L_\odot$, we can see that the larger values are far from the identity and this could be due to the impact of $T_\mathrm{eff}$ on our estimate.
However, our values in $T_\mathrm{eff}$ are in very good agreement with those found through Simbad for several articles for each target (see the plot at the top of Fig. \ref{teff_compSimbad} in appendix \ref{aa.literature}).

\subsection{Surface-brightness and color relation}

As in section \ref{ss.sb} we found that our SB does not depend on $\log g$ and $Z$, we conclude that this set of quiet stars is ideal for performing an SBCR calibration. However, some recent work on FGKM dwarfs \cite{Kiman24} and Cepheids \citep{Bailleul2026} show that the SBCR based on Gaia photometries seem to be quite sensitive to metallicity.

\begin{figure*}[htb!]
    \centering
    \includegraphics[width=0.33\textwidth]{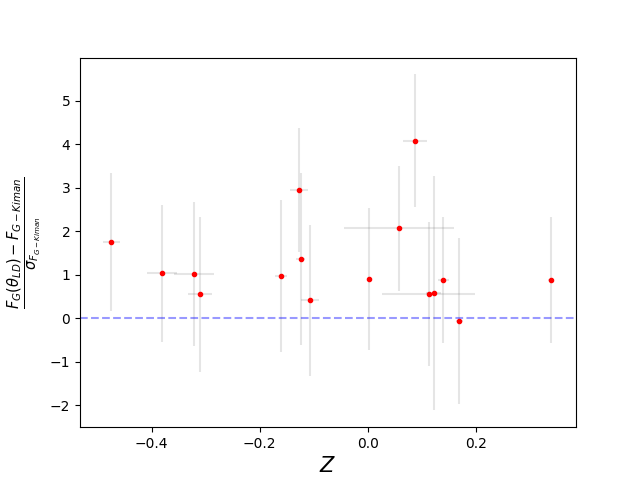}
    \includegraphics[width=0.33\textwidth]{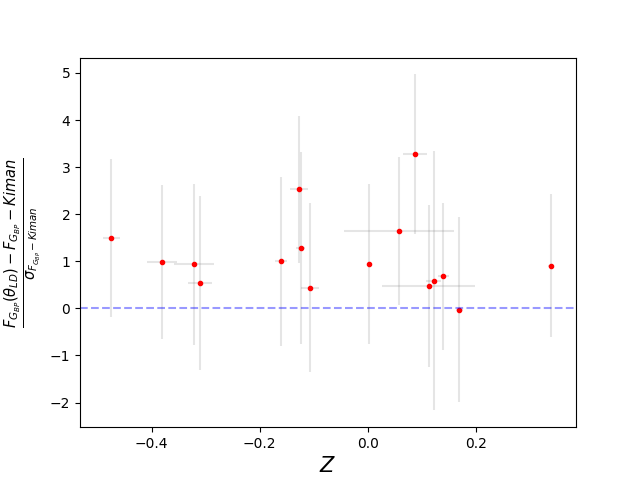}
    \includegraphics[width=0.33\textwidth]{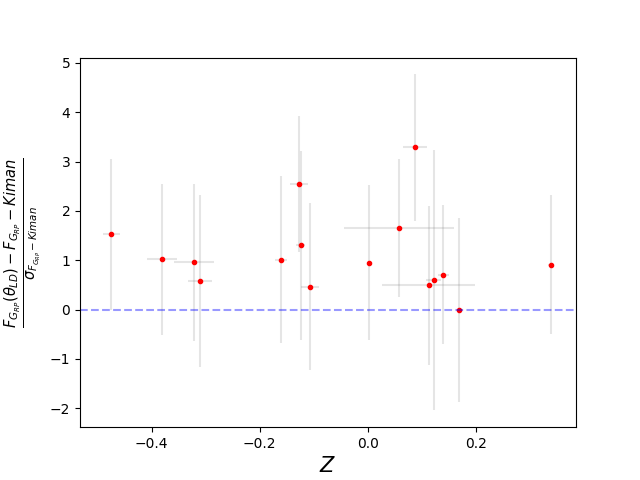}
    \caption{Difference between the surface brightness calculated in this work and that of \cite{Kiman24} weighted by the \citeauthor{Kiman24} sigma as a function of the metallicity, $Z$, estimated in this work. 
    }
    \label{comparisonSBCR}
\end{figure*}

In Fig. \ref{comparisonSBCR} we show the comparison of our results with those reported by \cite{Kiman24} where we have plotted the difference between the two of them weighted by the sigma of \cite{Kiman24} as a function of the $Z$ obtained in this work. 

From this figure, we can see that our data is not in good agreement with the relation found in the literature.
The difference between $F_\lambda$ estimated here and the one reported in \cite{Kiman24} is positive in the three $\lambda$-cases studied in this work and they are also independent of the metallicity. 
These discrepancies in the trend may be due to the fact that \cite{Kiman24} reported a non-linear SBC relationship with a very wide colour coverage across a very broad range of spectral types for FGKM dwarfs. 
As our objective is to study the SBC relation for the PLATO mission, we have only considered FGK stars, as explained in section \ref{ss.sample} and also following the suggestions in the works of \citealt{Salsi20, Salsi21}, which reported a clear difference in the SBCR for FGK and Ms stars, when considering dwarfs and subgiants. 
Another possibility to explain part of the difference between the non-linear SBCR from \cite{Kiman24} for FGK stars and the linear SBCR presented in this work for FGK, could be the methodology implemented for deriving the limb-darkened angular diameters. 
In this work, we apply a novel homogeneous and robust approach based on polychromatic interferometric observations while \cite{Kiman24} used various interferometric data available in the literature based on different approaches.

\begin{figure}[htb!]
    \centering
    \includegraphics[width=0.47\textwidth]{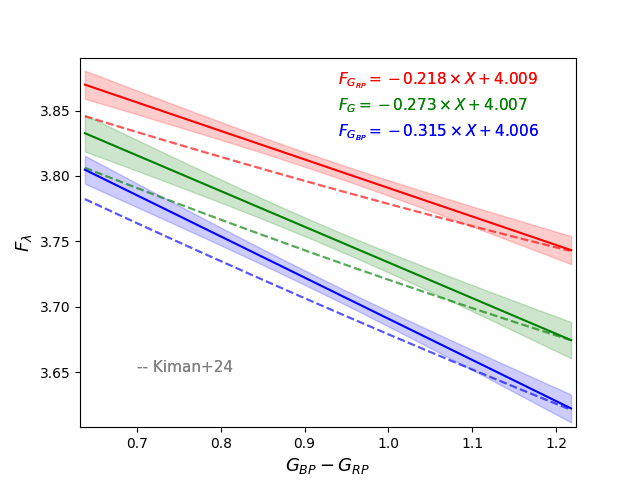}
    \caption{Difference between the surface brightness calculated in this work (solid lines) and the one from \citealt{Kiman24} (dashed lines) for the bands $G$ in green, $G_{BP}$ in blue and $G_{RP}$ in red. The shaded area corresponds to $1\sigma$ deviation for each calibration performed in this work.
    }
    \label{comparisonSBCR-Kiman}
\end{figure}

We then proceeded to perform our own calibration on the SBC relation.
The results are reported in Table \ref{tab_sbcrRes} and in Fig. \ref{comparisonSBCR-Kiman} we present the new calibration performed in this work and again the comparison with the one obtained in \cite{Kiman24} for the three Gaia bands, $F_G$, $F_{G_{BP}}$ and $F_{G_{RP}}$. 
Given that this is a homogeneous study both in terms of obtaining the fundamental parameters and analysing the SBC relationship, and that we did not find any dependence and/or relationship between our SBCR and $\log g$ or $Z$, it is important for the future to be able to analyse how stars in other ranges of $\log g$, $Z$ and $(G_{BP} - G_{RP})$ may affect the calibration carried out in this study.
On the other hand, it is important to analyse how star spots, multiplicity and environment can impact the SBCR calibrated from quiet stars such as those in the present work.

\begin{table}[htb!]
\caption{In this work, we have considered $F_\lambda = A \times X + B$, where $X = (G_{BP} - G_{RP})$ is the Gaia color index.}
\begin{center}
\label{tab_sbcrRes}    
\begin{tabular}{lccc}
\hline\hline\noalign{\smallskip}
$\lambda$     & A                  & B                 & $\sigma_{RMS}$ \\
\hline\noalign{\smallskip}

$G$      & -0.273 $\pm$ 0.018 & 4.007 $\pm$ 0.017 & 0.012          \\
$G_{BP}$ & -0.315 $\pm$ 0.013 & 4.006 $\pm$ 0.013 & 0.009          \\
$G_{RP}$ & -0.218 $\pm$ 0.014 & 4.009 $\pm$ 0.013 & 0.009 \\  
\hline\noalign{\smallskip}
\end{tabular}
\end{center}
\end{table}

\section{Conclusions} \label{s.conclusion}

In this work, we presented a homogeneous study of FGK stars within the framework of the PLATO space mission.
For the first time, we have combined three instruments operating in different bands to measure the angular diameters of stars and thus estimate their stellar radii with an average precision of 2.4\%.
Despite all the unforeseen events with the SPICA instrument, from 14 targets observed in \textit{R} band, we were able to measure 10 angular diameters with an average precision of 6\%. 
In addition, for the first time, we reached the second lobe using data from the SPICA instrument for the stars HD 3651, HD 30652, HD 19373, HD 142860, and HD 173667. 
For some targets, the second-lobe measurements allow the limb-darkening coefficient to be fitted independently, leading to more accurate and less model-dependent angular diameters.
This encourages us, then to move on to studying stars with angular diameters even smaller than those presented in this work.

On the other hand, we have calibrated the SBCR for quiet stars that are not affected by metallicity and/or surface gravity. This relationship is important to have as a reference point, given that improvements in the optical interferometer will allow us, in the future, to include more active stars (considering magnetic activity, star spots, multiplicity, and environment) to understand how different processes can affect or impact the relationship between surface brightness and color reported in this work.
Here is also important to highlight the necessity of a homogeneous and accurate photometric catalogue in the $K$-band in order to perform the SBCR calibration in order to minimize the dispersion in the relation.

Taking all this into account, the steps to be taken are to continue monitoring the stars presented in this study to improve observations in the optical range and thus enhance the precision of angular diameters in the \textit{R} band.
On the other hand, we will continue our analysis of K giant stars, which constitute a second goal, as mentioned in the introduction to this work, and to conduct a comparison between K dwarf/subgiant stars (this work) and giants and their impact on the SBCR relationship.

Furthermore, continuing to analyse stellar parameters in the most homogeneous way possible will allow us to further investigate how these can affect the SBCR relation and, therefore, the estimation of stellar radii according to, for example, the context of the PLATO space mission.

\begin{acknowledgements}
      This project has received funding from the European Research Council (ERC) under the European Union’s Horizon 2020 research and innovation programme (Grant agreement No. 101019653). SPICA has been funded by CNRS, Observatoire de la Côte d'Azur, Université Côte d'Azur, Région Sud, and the University of Aarhus. 
      This work is based upon observations obtained with the Georgia State University Center for High Angular Resolution Astronomy Array at Mount Wilson Observatory.  The CHARA Array is supported by the National Science Foundation under Grant No. AST-2034336 and AST-2407956. Institutional support has been provided from the GSU College of Arts and Sciences, Office of the Provost, and Office of the Vice President for Research and Economic Development.
      We warmly thank Chris Farrington, Becky Flores, Olli Majoinen, Heven Renteria, Norm Vargas, Cyril Pannetier and Cyprien Lanthermann for their excellent support during the night operations, and Victor Castillo, Larry Webster, and Craig Woods for all the support during the installation of SPICA, PSAUM, and also during our various stays at Mount Wilson.
      SK acknowledges support from an ERC Consolidator Grant (Grant Agreement ID 101003096). MIRC-X has been build with funds from an ERC Starting Grant (Grant Agreement No.\ 639889) and an STFC equipment grant ST/X005143/1. JDM acknowledges funding for the development of MIRC-X (NASA-XRP NNX16AD43G, NSF-AST 2009489) and MYSTIC (NSF-ATI 1506540, NSF-AST 1909165).
      The authors thank Grzegorz Pietrzynski, Dariusz Graczyk and  Mikolaj Kaluszynski for their help in setting the ISSP-SBCR objectives and samples connected to distance scale calibration.
      This research has made use of the SIMBAD database, CDS, Strasbourg Astronomical Observatory, France. 
      This research has made use of the VizieR catalogue access tool, CDS, Strasbourg Astronomical Observatory, France (DOI : 10.26093/cds/vizier).
      This research has made use of the Jean-Marie Mariotti Center \texttt{SearchCal} service\footnote{Available at \url{http://www.jmmc.fr/searchcal}}, which involves the JSDC and JMDC catalogues. 
      This research has made use of the Jean-Marie Mariotti Center \texttt{Aspro} service\footnote{Available at \url{http://www.jmmc.fr/aspro}}.
\end{acknowledgements}

%
%

\bibliographystyle{aa}
\small
\bibliography{biblio}

\begin{appendix} 
\section{Complementary data information} \label{aa.complementData}

In this section, we list the programme IDs for each of the public-domain spectra used in this study.
\newline

\noindent
\textbf{HARPS:} 075.D-0760(A), 075.C-0689(A), 078.C-0209(B), 076.C-0279(A), 076.C-0279(B), 076.C-0279(C), 077.C-0295(C), 086.D-0460(A), 110.24BB.001, 60.A-9709(G), 086.D-0460(A), 079.A-9017(A), 076.A-9006(A), 096.C-0053(A), 081.D-0870(A), 081.D-0531(A), 096.C-0708(A), 075.D-0194(A), 076.D-0130(A), 076.D-0130(B), 076.D-0130(C), 074.D-0131(B), 073.D-0038(A), 073.D-0038(C), 073.D-0038(D), 074.D-0131(C), 072.C-0096(A), 072.C-0096(B), 081.D-0065(B), 081.D-0065(c), 081.D-0065(D), 081.D-0065(E), 079.D-0075(C), 079.D-0075(D), 079.D-0075(E), 080.D-0086(A), 080.D-0086(B), 080.D-0086(C), 080.D-0086(D), 078.D-0071(A), 078.D-0071(B), 074.D-0131(A), 074.D-0131(B), 078.D-0071(C), 078.D-0071(D).

\noindent
\textbf{FEROS:} 60.A-9122(B), 087.D-0781(A), 0101.A-9012(A), 090.A-9003(A).

\noindent
\textbf{HARP-N:} CAT18A\_115, A32DDT2.

\noindent
\textbf{ELODIE:} 150.

\noindent
\textbf{SOPHIE: } 13B.PNP.DIFO, VSS, 06B.PNPS.FBOU, 11A.TECH.BOU, 07A.PNPS.FOSS, 07A.PNP.CONS, 14A.TECH.BOUC, 08A.PNP.CONS, 16A.PNP.COUR, 12A.PNP.MOUT, 14A.PNP.HEBR, 16B.PNP.HEBR, 18A.PNP.SANT.


\section{Comparison with the literature} \label{aa.literature}

\begin{figure}[htb!]
    \centering
    \includegraphics[width=0.47\textwidth]{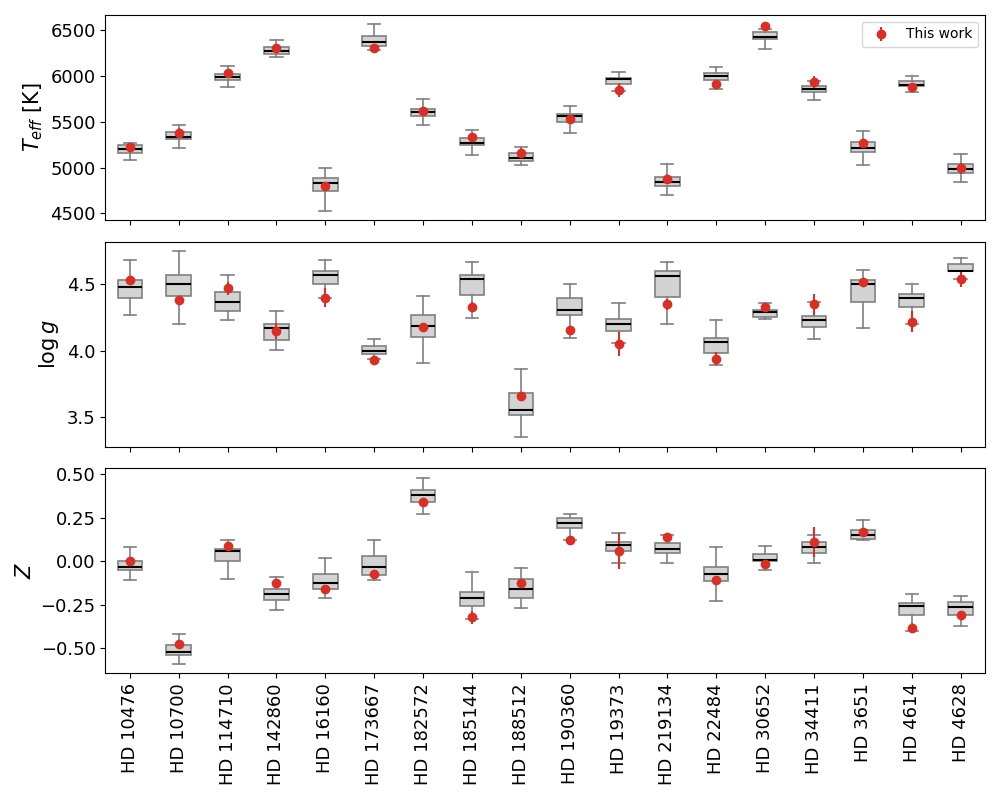}
    \caption{Box-plots showing the distribution of literature values (post-2000) extracted from Simbad for each star. From top to bottom: effective temperature ($T_\mathrm{eff}$), surface gravity ($\log g$), and metallicity ($Z$). Box-plots provide a visual summary of the literature spread, including median and interquartile range. Red markers indicate the measurements obtained in this work, with error bars representing the uncertainties.}
    \label{teff_compSimbad}
\end{figure}

\newpage

\section{Visibilities} \label{aa.vis}

\begin{figure}[htb!]
    \centering
    \includegraphics[width=0.45\textwidth]{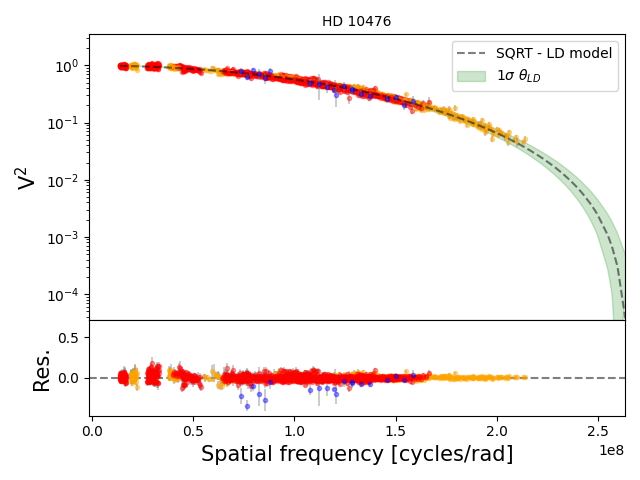}
    \includegraphics[width=0.45\textwidth]{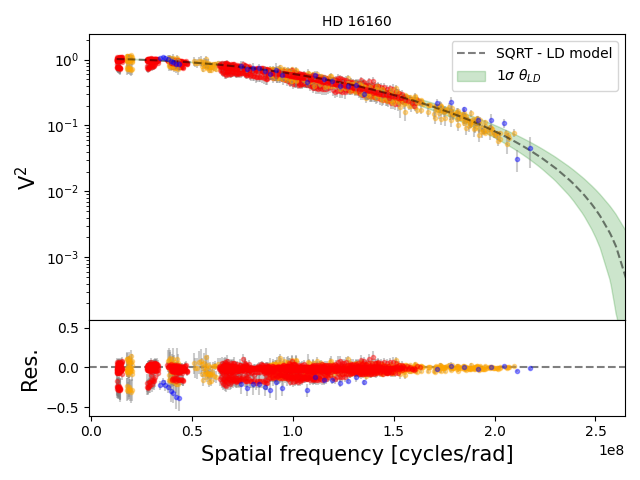}
    \includegraphics[width=0.45\textwidth]{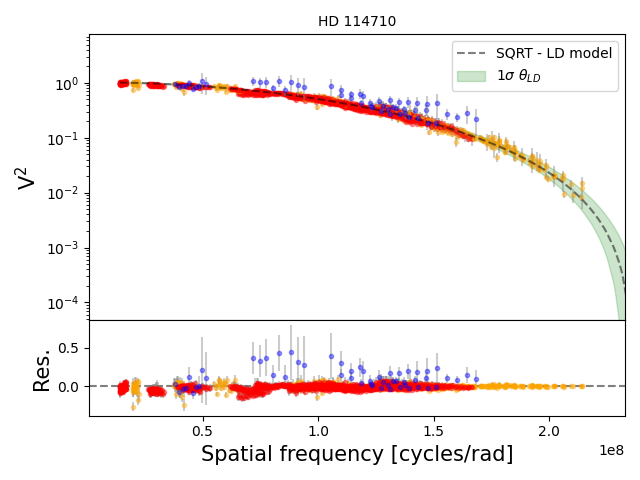}

    \caption{Squared visibilities for the stars in our sample. We have represented SPICA, MIRCX and MYSTIC data in blue, orange and red points, respectively. The dashed lines represents the SQRT-LD angular diameter plus the background  model considered in this work and the green shadow represents the $1\sigma$ deviation. }
    \label{visibilities}
\end{figure}

\begin{figure*}[htb!]
    \centering
    \ContinuedFloat
    \captionsetup{list=off,format=cont}
    \includegraphics[width=0.45\textwidth]{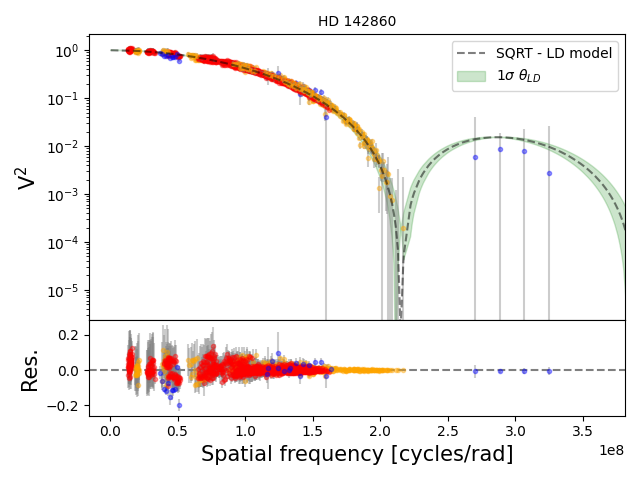}
    \includegraphics[width=0.45\textwidth]{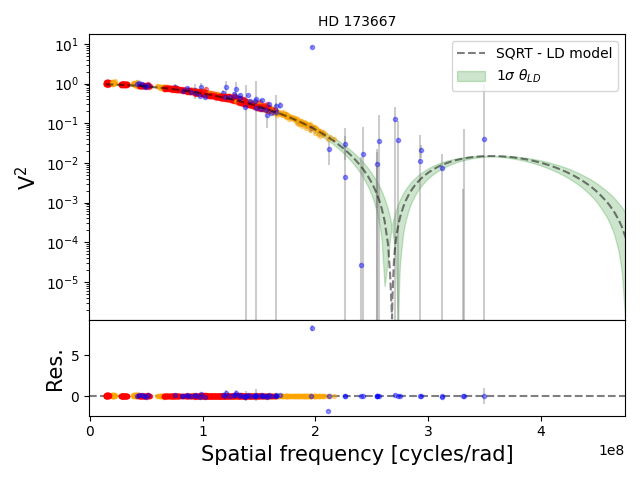}
    \includegraphics[width=0.45\textwidth]{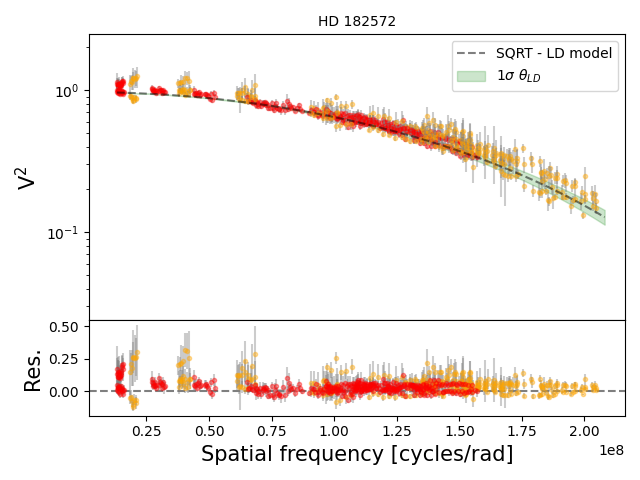}
    \includegraphics[width=0.45\textwidth]{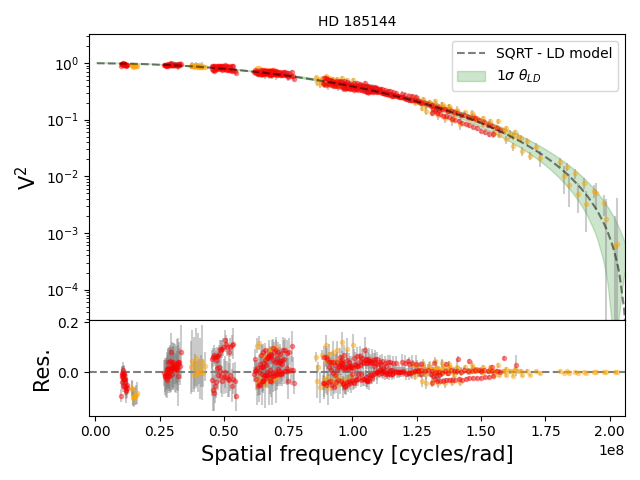}
    \includegraphics[width=0.45\textwidth]{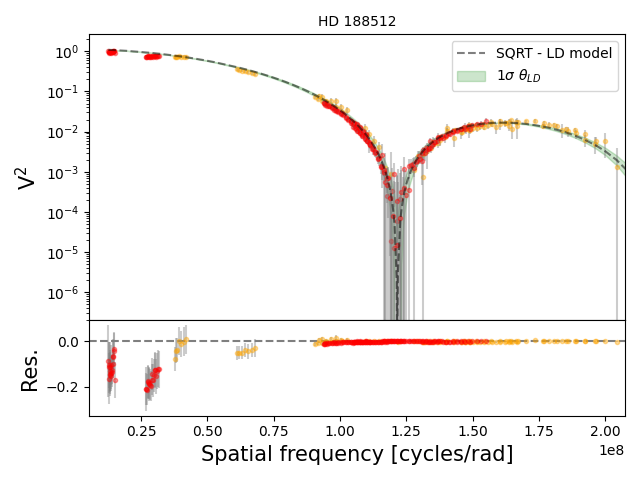}
    \includegraphics[width=0.45\textwidth]{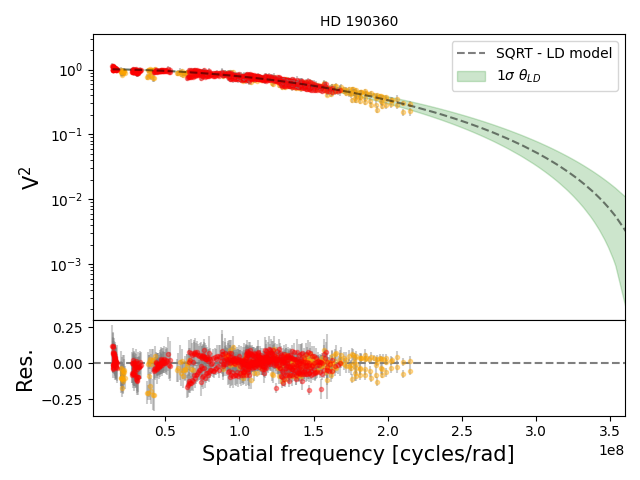}
    \includegraphics[width=0.45\textwidth]{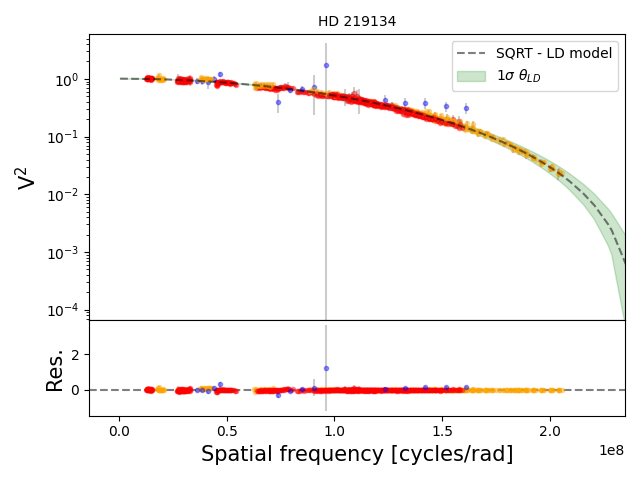}
    \includegraphics[width=0.45\textwidth]{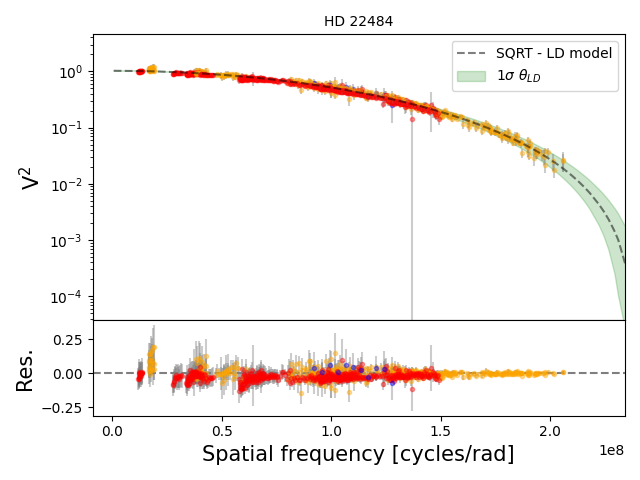}

    \caption{ }
    \label{visibilities}
\end{figure*}

\begin{figure*}[htb!]
    \centering
    \ContinuedFloat
    \captionsetup{list=off,format=cont}
    
    \includegraphics[width=0.45\textwidth]{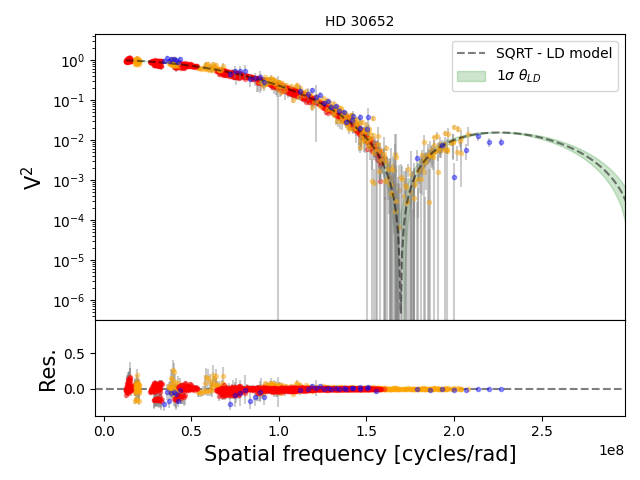}
    \includegraphics[width=0.45\textwidth]{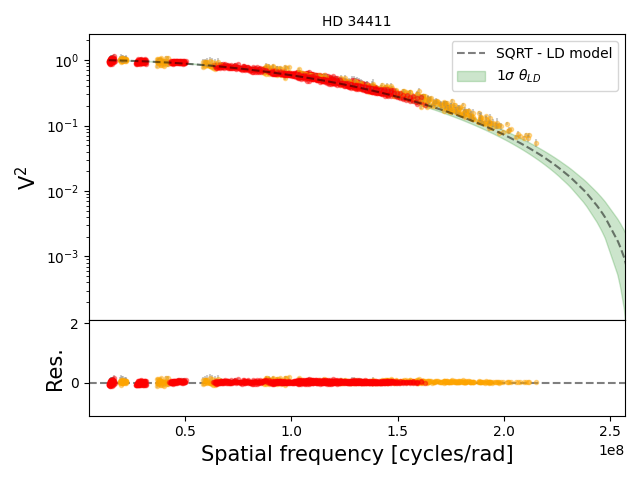}
    \includegraphics[width=0.45\textwidth]{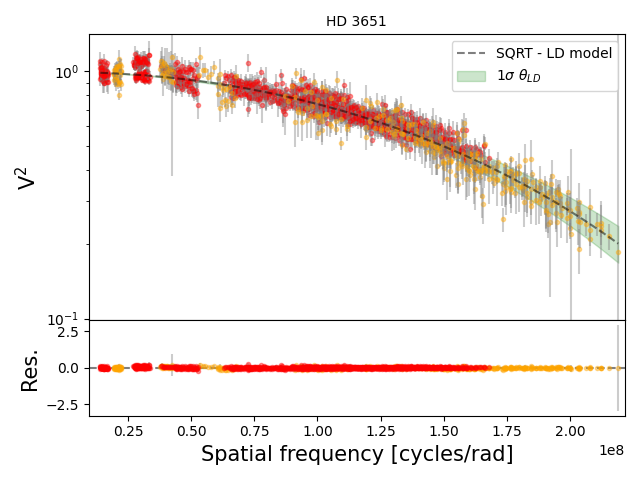}
    \includegraphics[width=0.45\textwidth]{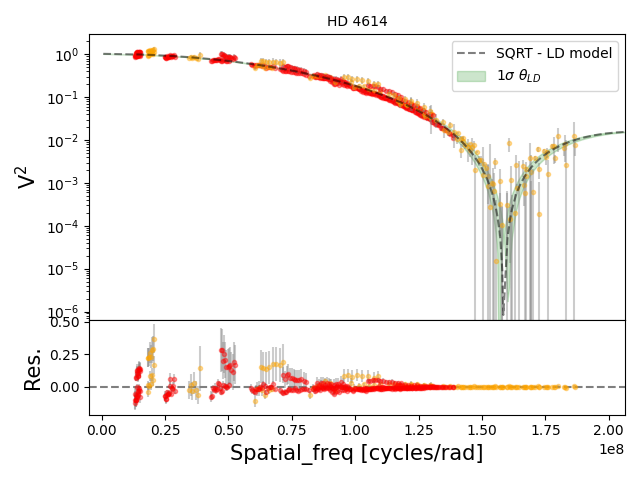}
    \includegraphics[width=0.45\textwidth]{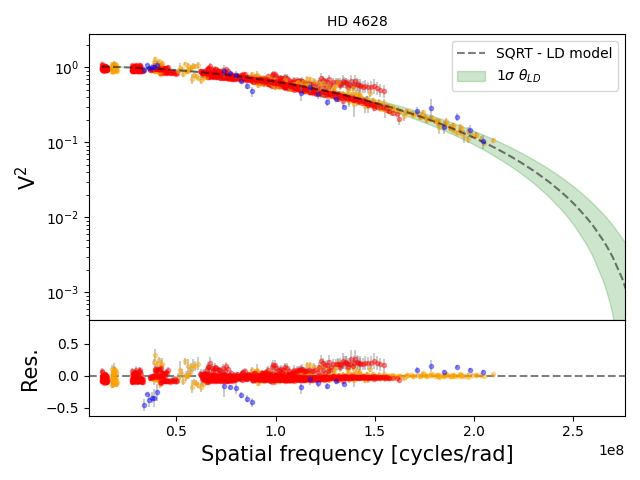}

    \caption{ }
    \label{visibilities}
\end{figure*}

\end{appendix}

\end{document}